\documentclass[12pt]{article}
\usepackage[dvipdfmx]{graphicx}
\usepackage{amssymb, amsmath, bm, float, epsfig, color, slashed, dsfont, subcaption, multirow, arydshln}
\usepackage[centering]{geometry}
\usepackage{cite}

\allowdisplaybreaks[4]

\newcommand{\red}[1]{\textcolor{red}{#1}}

\makeatletter
\@addtoreset{equation}{section}
\makeatother

\begin{document}

\thispagestyle{empty}

\begin{flushright}
KOBE-TH-22-04 \\
\end{flushright}
\vspace{40pt}
\begin{center}
{\Large\bf{Index and winding numbers on $T^2/\mathbb{Z}_N$ orbifolds with magnetic flux}} \\

\vspace{40pt}
{\bf{Hiroki Imai$^\dagger\hspace{.5pt}$}\footnote{E-mail: 216s103s@stu.kobe-u.ac.jp}}, \, 
{\bf{Makoto Sakamoto$^\dagger\hspace{.5pt}$}\footnote{E-mail: dragon@kobe-u.ac.jp}}, \, 
{\bf{Maki Takeuchi$^\dagger\hspace{.5pt}$}\footnote{E-mail: 191s107s@stu.kobe-u.ac.jp}}, \, 
{\bf{Yoshiyuki Tatsuta$^\ast\hspace{.5pt}$}\footnote{E-mail: yoshiyuki.tatsuta@sns.it }} \\

\vspace{40pt}
{\it $^{\dagger}$ Department of Physics, Kobe University, Kobe 657-8501, Japan \\[5pt]
$^{\ast}$ Scuola Normale Superiore and INFN, Piazza dei Cavalieri 7, 56126 Pisa, Italy} \\
\end{center}
\vspace{30pt}
\begin{abstract}
	\noindent
We analyze the number of independent chiral zero modes and the winding numbers at the fixed points on $T^2/{\mathbb{Z}}_N$ ($N=2,3,4,6$) orbifolds with magnetic flux. In the case of $N=2$, we derive the index formula $n_{+}-n_{-}=M/2+(-V_{+}+V_{-})/4=M/2-V_{+}/2+1$ by using the trace formula, where $n_{\pm}$ are the numbers of the $\pm$ chiral zero modes and $V_{\pm}$ are the sums of the winding numbers at the fixed points on $T^2/{\mathbb{Z}}_2$. We also  obtain the formula $n_{+}-n_{-}=M/N+(-V_{+}+V_{-})/(2N)=M/N-V_{+}/N+1$ for $N=3,4,6$ {under} an assumption.
\end{abstract}

\newpage
\setcounter{page}{2}
\setcounter{footnote}{0}

\section{Introduction}
In higher dimensions, various works have been done to explain the mysteries of the Standard Model which are the generation problem of quarks and leptons~\cite{Sakamoto:2020pev,Abe:2008sx,Abe:2015yva,Libanov:2000uf,Frere:2000dc,PhysRevD.65.044004,PhysRevD.73.085007,Gogberashvili:2007gg,Guo:2008ia,PhysRevLett.108.181807}, mass hierarchy~\cite{Cremades:2004wa,Abe:2008sx,Arkani-Hamed:1999ylh,Dvali:2000ha,Gherghetta:2000qt,Kaplan:2000av,Huber:2000ie,Kaplan:2001ga,Fujimoto:2012wv,PhysRevD.97.115039,PhysRevD.90.105006}, CP violation~\cite{PhysRevD.88.115007,PhysRevD.90.105006,Kobayashi:2016qag,Buchmuller:2017vho,PhysRevD.97.075019}. Especially, string theory is a strong candidate beyond the Standard Model. However, a crucial difficulty is {to obtain} the chiral spectrum. The method to obtain the chiral spectrum is known as toroidal orbifold compactification~\cite{Dixon:1985jw,Dixon:1986jc}, as a simple model. 

The Atiyah-Singer index theorem~\cite{Atiyah:1963zz} is a powerful tool to obtain the number of chiral fermions, and states that the 
index of a Dirac operator ${\slashed{D}}$ 
\begin{align}
{\rm{Ind}}(i\slashed D)\equiv n_{+}-n_{-}
\label{ASindex}
\end{align}
is topologically invariant. Here, $n_{\pm}$ are the numbers of ${\pm}$ chiral zero modes for the Dirac operator. For example, the Atiyah-Singer index theorem on the torus $T^2$ with magnetic flux is known as~\cite{Witten:1984dg,Green:1987mn}
\begin{align}
n_{+}-n_{-}=\frac{q}{2\pi}\int_{T^2}F=M,
\label{Indextorus}
\end{align}
where $M$ denotes the flux quanta in the torus compactification. The number of chiral zero modes is decided by the flux quanta. In the torus model, there is only one model that obtains three-generation: the case of $M=3$.

On the other hand, in the $T^2/{\mathbb{Z}}_N\,\,(N=2,3,4,6)$ orbifolds with magnetic flux background, it is known that there are a lot of models to obtain the three-generation~\cite{Sakamoto:2020pev,Abe:2013bca,Abe:2014noa,PhysRevD.96.096011}. In the previous paper \cite{Sakamoto:2020pev}, a complete list of the number of chiral zero modes was found and the index $n_{+}-n_{-}$ of the models was given by a simple formula
\begin{align}
n_{+}-n_{-}=\frac{M-V_{+}}{N}+1,
\label{Zeromodecountingformula}
\end{align}
where $V_{+}$ is the sum of the winding numbers for $+$ chirality modes at the fixed points of the $T^2/{\mathbb{Z}}_N\,\,(N=2,3,4,6)$ orbifolds. We call this formula the ``zero-mode counting formula". It is not, however, clear whether this formula can be regarded as the index theorem on the $T^2/{\mathbb{Z}}_N$ orbifolds with magnetic flux. This is because in Ref.\cite{Sakamoto:2020pev} $n_{+}-n_{-}$ and $\frac{M-V_{+}}{N}+1$ have been computed separately, and verified the equality of them by comparing the values for each case of $M$, the ${\mathbb{Z}}_N$ eigenvalue, the Scherk-Schwarz (SS) twist phase $(\alpha_1,\alpha_2)$ and $N$.

In the paper~\cite{Sakamoto:2020vdy}, it was derived the index theorem on the $T^2/{\mathbb{Z}}_N$ orbifolds without magnetic flux:
\begin{align}
n_{+}-n_{-}=-\frac{V_{+}}{N}+1=\frac{1}{2N}(-V_{+}+V_{-}),
\label{indexwithoutM}
\end{align}
where $V_{-}$ is the sum of the winding numbers for $-$ chirality modes at the fixed points with {a} relation $V_{+}+V_{-}=2N$. In this work, the terms $-\frac{V_{+}}{N}+1$ have been derived by use of the trace formula without magnetic flux.
{Since there is no magnetic flux, there is no contribution from the bulk, and the index is represented by the contribution at the fixed points on the $T^2/{\mathbb{Z}}_N$ orbifolds.}

{In} Eq.\eqref{Zeromodecountingformula}, $M/N$ can be considered as the flux contribution of the bulk as in Eq.\eqref{Indextorus}. {Another} contribution at the fixed points $-\frac{V_{+}}{N}+1$ is consistent with Eq.\eqref{indexwithoutM}, which was derived as the index theorem. Therefore, Eq.\eqref{Zeromodecountingformula} is expected to be derived as the index theorem. In this paper, in order to clarify the relationship between Eq.\eqref{Zeromodecountingformula} and the index theorem, we will check whether the right-hand side of Eq.\eqref{Zeromodecountingformula} is derived {as the index theorem}.
We will compute the index $n_+-n_-$ on the $T^2/{\mathbb{Z}}_N$ orbifolds with magnetic flux by use of the trace formula
\begin{align}
{\rm{Ind}}(i\slashed{D})_{\eta}=\lim_{\rho \to \infty} {\rm{tr}}[\sigma_3 e^{\slashed{D}^2/\rho^2}]_{\eta}.
\label{traceformula}
\end{align}
and rederive the formula
\begin{align}
n_{+}-n_{-}=\frac{M-V_{+}}{N}+1=\frac{M}{N}+\frac{1}{2N}(-V_{+}+V_{-}).
\label{IndexZNintro}
\end{align}
with {a} relation $V_{+}+V_{-}=2N$.
The proof is the main result of this paper.

This paper is organized as follows. In Section 2, we review the mode functions on the torus. In Section 3, we construct the mode functions on the $T^2/{\mathbb{Z}}_2$ orbifold with magnetic flux.  In Section 4, we evaluate the trace formula \eqref{traceformula} by using the complete set of the mode functions on the $T^2/{\mathbb{Z}}_2$ orbifold. In Section 5, we also evaluate the trace formula \eqref{traceformula} on the $T^2/{\mathbb{Z}}_N\,\,(N=3,4,6)$ orbifolds {under} an assumption. In Section 6, we rewrite the results obtained in Sections 4 and 5 by winding numbers at the fixed points and get the index formula \eqref{IndexZNintro}. Section 7 is devoted to the discussion and conclusion. In Appendices, we mention our notation and also derive formulae used in our discussions.

\section{Mode functions on $T^2$}
In this section, we briefly review mode functions on the two-dimensional (2d) torus $T^2$ with magnetic flux background~\cite{Cremades:2004wa,Abe:2013bca,Abe:2014noa}.

\subsection{Setup}
In this paper, we consider a six-dimensional ($6$d) abelian gauge theory compactified on $T^2$. By use of the complex coordinate $z=y_1+y_2\tau$, where ${\bm{y}}=(y_1,y_2)\,\,(0\leq y_1,y_2<1)$ is the oblique coordinate, the torus $T^2$ is defined by the identification
\begin{align}
 z\sim z+1\sim z+\tau \qquad(\,\tau\in {\mathbb{C}}, \,\, {\rm{Im}}\tau >0\,)
 \label{torusid}
 \end{align}
under torus lattice shifts.

The non-zero magnetic flux $f$ on $T^2$ is obtained as $f=\int_{T^2} F$ with the field strength
\begin{align}
 F(z)=\frac{if}{2{\rm{Im}}\tau}dz \wedge d\bar{z}.
\end{align}
For $F=dA$, the (1-form) vector potential is given by
\begin{align}
 A(z)=\frac{f}{2{\rm{Im}}\tau}{\rm{Im}}(\bar{z} dz).
 \end{align}
The torus lattice shifts on the vector potential should be accompanied by the gauge transformations as
 \begin{align}
& A(z+1)=A(z)+d\Lambda_1(z),\\
& A(z+\tau)=A(z)+d\Lambda_2(z),
 \end{align}
where $\Lambda_1(z)$ and $\Lambda_2(z)$ are gauge parameters given by
\begin{align}
\Lambda_1(z)=\frac{f}{2{\rm{Im}}\tau}{\rm{Im}}z, \qquad
\Lambda_2(z)=\frac{f}{2{\rm{Im}}\tau}{\rm{Im}}\bar{\tau} z.
\end{align}

We consider a 6d Weyl fermion in magnetic flux background:
\begin{align}
{\mathcal{L}}_{{\rm{6d}}}=i\bar{\Psi}\Gamma^{K} D_{K} \Psi, \qquad \Gamma_7 \Psi = \Psi,
 \end{align}
 where $K(=0,1,2,3,5,6)$ is the 6d spacetime index and $D_K=\partial_K-iqA_K$ is the covariant derivative. $\Gamma^{K}$ denote 6d Gamma matrices and $\Gamma_{7}$ is the 6d chiral operator. 
 
The 6d Weyl fermion $\Psi(x,z)$ can be decomposed into 4d Weyl right/left-handed fermions 
$\psi_{{\rm{R}}/{\rm{L}}}^{(4)}(x)$ as 
\begin{align}
	\Psi(x, z) = \sum_{n, \hspace{.5pt} j} 
	\bigl\{ \psi^{(4)}_{\textrm{R}, \hspace{.5pt} n, \hspace{.5pt} j}(x) \otimes \psi^{(2)}_{+, \hspace{.5pt} n, \hspace{.5pt} j}(z) + \psi^{(4)}_{\textrm{L}, \hspace{.5pt} n, \hspace{.5pt} j}(x) \otimes \psi^{(2)}_{-, \hspace{.5pt} n, \hspace{.5pt} j}(z) \bigr\},
\end{align}
 where $x^{\mu} \,\,(\mu=0,1,2,3)$ denotes the 4d Minkowski coordinate. The 4d Weyl fermions $\psi_{{\rm{R}}/{\rm{L}},n,j}^{(4)}(x)$ satisfy $ \gamma_5 \psi^{(4)}_{{\rm{R}/{L}},n,j}(x)=\pm \psi^{(4)}_{{\rm{R}/{L}},n,j}(x)$. The 2d Weyl fermions $\psi_{\pm,n,j}^{(2)}(z)$ are expressed as
  \begin{gather}
	\psi^{(2)}_{+, \hspace{.5pt} n, \hspace{.5pt} j} (z)= 
	\begin{pmatrix}
		\psi_{+, \hspace{.5pt} n}^j (z)\\[3pt] 0
	\end{pmatrix}
	, \qquad \psi^{(2)}_{-, \hspace{.5pt} n, \hspace{.5pt} j}(z)= 
	\begin{pmatrix}
		0 \\[3pt] \psi_{-, \hspace{.5pt} n}^j(z)
	\end{pmatrix},
	\label{psipm}
\end{gather}
where $n$ and $j$ label the Landau level and the degeneracy of mode functions on each level, respectively.

Due to the existence of the vector potential, the 2d Weyl fermions have to obey the pseudo periodic boundary conditions
\begin{align}
\psi_{\pm, \hspace{.5pt} n}^j(z + 1) = U_1(z) \psi_{\pm, \hspace{.5pt} n}^j(z), \qquad \psi_{\pm, \hspace{.5pt} n}^j(z + \tau) = U_2(z) \psi_{\pm, \hspace{.5pt} n}^j(z)
\label{BCs}
\end{align}
with
\begin{align}
U_i(z) = e^{i q \Lambda_i (z)} e^{2\pi i \alpha_i} \quad (i=1,2),
\label{BCss}
\end{align}
where $\alpha_i \,\, (i=1,2)$ corresponds to the Scherk-Schwarz twist phase. The consistency condition that the 2d Weyl fermions can be well defined on the torus leads to the magnetic flux quantization
\begin{align}
\frac{qf}{2\pi}\equiv M \in \mathbb{Z}.
\end{align}

The mode functions $\psi_{\pm, \hspace{.5pt} n}^j(z)$ in Eq.\eqref{psipm} are required to satisfy the equations
\begin{align}
-2D_z \psi_{-, \hspace{.5pt} n}^j(z)&=m_n \psi_{+, \hspace{.5pt} n}^j(z),\label{MD1} \\
2D_{\bar{z}} \psi_{+, \hspace{.5pt} n}^j(z)&=m_n \psi_{-, \hspace{.5pt} n}^j(z),
\label{MD2}
\end{align}
where
\begin{align}
D_z \equiv \partial_z -\frac{\pi M}{2{\rm{Im}}\tau}\bar{z},&\qquad 
D_{\bar{z}}\equiv \partial_{\bar{z}} +\frac{\pi M}{2{\rm{Im}}\tau}{z}
\end{align}
with the orthonormality condition
\begin{align}
\int_{T^2} dz d\bar{z}\, [\psi_{\pm,n}^{j}(z)]^{\ast}\psi_{\pm,m}^{k}(z)=\delta^{jk}\delta_{nm}.
\label{OBC}
\end{align}
Our interest is how many zero mode solutions exist for $m_n=0$. In the next subsection, we discuss zero modes on $T^2$.

\subsection{Zero mode functions on $T^2$}
It follows from Eqs.\eqref{MD1} and \eqref{MD2} that the zero mode functions $\psi_{\pm, \hspace{.5pt} 0}^j(z)$ should satisfy
 \begin{align}
 D_{\bar{z}} \psi_{+, \hspace{.5pt} 0}^j(z)=\left(\partial_{\bar{z}} +\frac{\pi M}{2{\rm{Im}}\tau}{z}\right)\psi_{+, \hspace{.5pt} 0}^j(z)=0,\label{Zero1}\\
D_z \psi_{-, \hspace{.5pt} 0}^j(z)=\left(\partial_z -\frac{\pi M}{2{\rm{Im}}\tau}\bar{z}\right)\psi_{-, \hspace{.5pt} 0}^j(z)=0.
\label{Zero2}
\end{align}
In the case of $M>0$, the zero mode solutions that satisfy Eq.\eqref{Zero1} and the boundary conditions \eqref{BCs} are found to be
 \begin{align}
 \psi_{T^2+,0}^{(j+\alpha_1,\alpha_2)}(z) &= \mathcal{N}_{T^2}\,e^{i \pi M z \, {\rm Im}\,z / {\rm Im}\, \tau} \, \vartheta
\begin{bmatrix}
\tfrac{j + \alpha_1}M \\[3pt] -\alpha_2
\end{bmatrix}
(Mz, M\tau),
\label{toruses}
\end{align}
where $j=0,1,\cdots,M-1$. ${\cal{N}}_{T^2}$ is  the normalization constant and is given by\footnote{{In \cite{Cremades:2004wa}, the normalization factor is
\begin{align}
{\cal{N}}_{T^2}=\left(\frac{2M{\rm{Im}\tau}}{\mathcal{A}^2}\right)^{\frac{1}{4}},
\end{align}
where $\mathcal{A}={\rm{Im}\tau}$ in this paper.}}
\begin{align}
{\cal{N}}_{T^2}=\left(\frac{2M}{\rm{Im}\tau}\right)^{\frac{1}{4}}.
\end{align}
$\vartheta$ is the theta function defined by
\begin{align}
\vartheta
\begin{bmatrix}
a \\[3pt]b
\end{bmatrix}
({z}, {\tau})
=\sum_{l=-\infty}^{\infty}e^{i\pi(a+l)^2 {\tau}} e^{2\pi i(a+l)({z}+b)}.
\end{align}
We note that there is no normalizable solution to Eq.\eqref{Zero2} for $M>0$.
 
We restrict our considerations to $M>0$ in this paper, although we can analyze the case of $M<0$ in a similar way. We emphasize that the number of the zero modes on $T^2$ with magnetic flux is $|M|$. Therefore, we obtain $|M|$-generation 4d chiral fermions.

\subsection{Kaluza-Klein mode functions on $T^2$}
In this subsection, we discuss Kaluza-Klein mode functions on $T^2$, which will be used in the proof of the index theorem on the 2d toroidal orbifolds $T^2/{\mathbb{Z}}_N\,(N=2,3,4,6)$.

From Eqs.\eqref{MD1} and \eqref{MD2}, the mode functions $\psi_{\pm, \hspace{.5pt} n}^j(z)$ obey
\begin{align}
\begin{pmatrix}
-4D_z D_{\bar{z}} & 0\\
0 &-4 D_{\bar{z}}D_z 
\end{pmatrix}
\begin{pmatrix}
\psi_{+, \hspace{.5pt} n}^j(z)\\
\psi_{-, \hspace{.5pt} n}^j(z)
\end{pmatrix}
=m_n^2
\begin{pmatrix}
\psi_{+, \hspace{.5pt} n}^j(z)\\
\psi_{-, \hspace{.5pt} n}^j(z)
\end{pmatrix}.
\label{Torusev}
\end{align}
The eigenvalue equation \eqref{Torusev} can easily be solved by introducing the annihilation and the creation operators as 
\begin{align}
&\hat{a}_{+} \equiv i\sqrt{\frac{{\rm{Im}}\tau}{\pi M}}D_{\bar{z}}=i\sqrt{\frac{{\rm{Im}}\tau}{\pi M}}
\left(\partial_{\bar{z}} +\frac{\pi M}{2{\rm{Im}}\tau}{z}\right),
\\
&\hat{a}_{+}^{\dagger} \equiv i\sqrt{\frac{{\rm{Im}}\tau}{\pi M}}D_z=i\sqrt{\frac{{\rm{Im}}\tau}{\pi M}}
\left(\partial_{{z}} -\frac{\pi M}{2{\rm{Im}}\tau}{\bar{z}}\right)
\end{align}
with
\begin{align}
&[\hat{a}_{+},\hat{a}_{+}^{\dagger}]=1.
\end{align}

\begin{figure}[!t]
\centering
\includegraphics[width=0.6\textwidth]{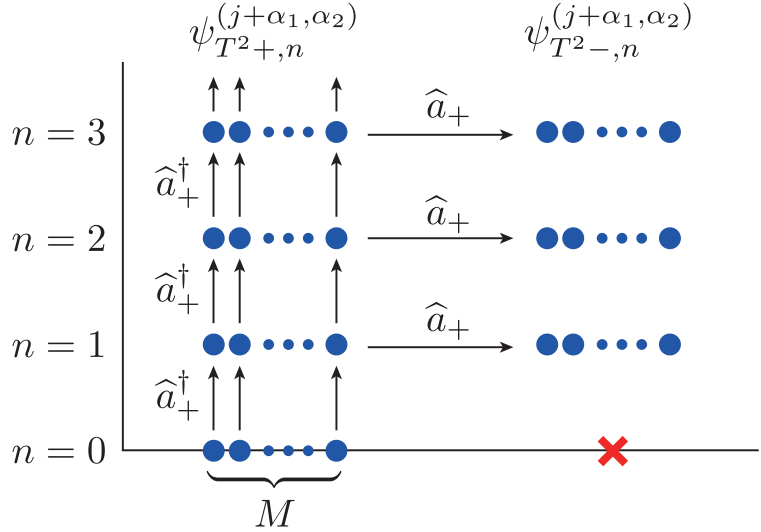}
\caption{The mass spectrum of $\psi_{T^2\pm,n}^{(j+\alpha_1,\alpha_2)}(z)$ for $M>0$. There are $M$ degenerate zero mode solutions $\psi_{T^2+,0}^{(j+\alpha_1,\alpha_2)}(z)$. The blue-filled circles correspond to zero modes and their Kaluza-Klein modes. 
The arrows mean that the $(n+1)$th modes $\psi_{T^2+,n+1}^{(j+\alpha_1,\alpha_2)}(z)$ can be obtained by acting the creation operator $\hat{a}_{+}^{\dagger}$ on $n$th modes $\psi_{T^2+,n}^{(j+\alpha_1,\alpha_2)}(z)$, and that the $n$th modes $\psi_{T^2-,n}^{(j+\alpha_1,\alpha_2)}(z)$ can be obtained by acting the annihilation operator $\hat{a}_{+}$ on the $n$th modes $\psi_{T^2+,n}^{(j+\alpha_1,\alpha_2)}(z)$.}
\label{figureT2}
\end{figure}

In terms of the creation operator $\hat{a}_{+}^{\dagger}$ and the zero mode functions $\psi_{T^2+,0}^{(j+\alpha_1,\alpha_2)}(z)$, the massive mode functions $\psi_{T^2{{\pm}},n}^{(j+\alpha_1,\alpha_2)}(z)$ can be constructed as
\begin{align}
\psi_{T^2+,n}^{(j+\alpha_1,\alpha_2)}(z) &\equiv  \frac{1}{\sqrt{n!}}(\hat{a}_{+}^{\dagger})^n\psi_{T^2+,0}^{(j+\alpha_1,\alpha_2)}(z), \\
\psi_{T^2-,n}^{(j+\alpha_1,\alpha_2)}(z)&=\frac{2}{m_n}D_{\bar{z}}\psi_{T^2+,n}^{(j+\alpha_1,\alpha_2)}(z) \notag \\
&=-i\psi_{T^2+,n-1}^{(j+\alpha_1,\alpha_2)}(z)\qquad (n=1,2,\cdots),
\label{KKmode-}
\end{align}
where we have used the relation \eqref{MD2} in the first equality of Eq.\eqref{KKmode-} and also used the mass eigenvalue
\begin{align}
m_n=\sqrt{\frac{4\pi M}{{\rm{Im}}\tau}n} \qquad (n=0,1,2,\cdots).
\end{align}
It should be emphasized that there is no zero mode function for $\psi_{-,0}$ with $M>0$, but there exist the non-zero mode functions for both $\psi_{T^2+,n}^{(j+\alpha_1,\alpha_2)}(z)$ and $\psi_{T^2-,n}^{(j+\alpha_1,\alpha_2)}(z)\,\,(n=1,2,\cdots\,\,$and$\,\, j=0,1,\cdots,|M|-1)$. The mass spectrum of $\psi_{T^2\pm,n}^{(j+\alpha_1,\alpha_2)}(z)$ is depicted in Figure \ref{figureT2}.

\section{Mode functions on $T^2/{\mathbb{Z}}_2$ orbifold}
In this section, we construct mode functions on the $T^2/{\mathbb{Z}}_2$ orbifold with magnetic flux. As we will see later, they are used to derive the index theorem on the $T^2/{\mathbb{Z}}_2$ orbifold.

\subsection{Zero mode functions on $T^2/{\mathbb{Z}}_2$}
In this subsection, we derive ${\mathbb{Z}}_2$ zero mode eigenstates on the $T^2/{\mathbb{Z}}_2$ orbifold. The $T^2/{\mathbb{Z}}_2$ orbifold is defined by the torus identification \eqref{torusid} and an additional ${\mathbb{Z}}_2$ one
\begin{align}
z \sim -z.
\end{align}
In the case of $T^2/{\mathbb{Z}}_2$, there is no restriction on $\tau$ except for ${\rm{Im}}\tau>0$. We often use $\omega=e^{i\pi}=-1$ for the $T^2/{\mathbb{Z}}_2$ orbifold $(\omega=e^{i2\pi/N}$ for the $T^2/{\mathbb{Z}}_N\,\,(N=3,4,6)$ orbifold). To be  consistent with the ${\mathbb{Z}}_2$ orbifold identification, the SS phase $(\alpha_1,\alpha_2)$ has to be quantized as~\cite{Abe:2013bca}
\begin{align}
 (\alpha_1,\alpha_2)=(0,0),(1/2,0),(0,1/2),(1/2,1/2).
 \end{align}

The ${\mathbb{Z}}_2$ zero mode eigenstates $\psi_{T^2/{\mathbb{Z}}_2+,0}^{(j+\alpha_1,\alpha_2)}(z)_{\eta}$ on $T^2/{\mathbb{Z}}_2$ satisfy
\begin{align}
\psi_{T^2/{\mathbb{Z}}_2+,0}^{(j+\alpha_1,\alpha_2)}(-z)_{\eta}=\eta\psi_{T^2/{\mathbb{Z}}_2+,0}^{(j+\alpha_1,\alpha_2)}(z)_{\eta},
\label{Z2ev}
\end{align}
 where $\eta=\pm1$ are ${\mathbb{Z}}_2$ eigenvalues. In terms of the zero mode functions $\psi_{T^2+,0}^{(j+\alpha_1,\alpha_2)}(z)$ on $T^2$, the ${\mathbb{Z}}_2$ eigenstates $\psi_{T^2/{\mathbb{Z}}_2+,0}^{(j+\alpha_1,\alpha_2)}(z)_{\eta}$ on $T^2/{\mathbb{Z}}_2$ can be constructed as 
 \begin{align}
\psi_{T^2/{\mathbb{Z}}_2+,0}^{(j+\alpha_1,\alpha_2)}(z)_{\pm1}\equiv{\cal{N}}_{+,\pm1}^{(j+\alpha_1,\alpha_2)}
[ \psi^{(j+\alpha_1,\alpha_2)}_{T^2+,0}(z)\pm \psi^{(j+\alpha_1,\alpha_2)}_{T^2+,0}(-z)],
\label{Z2es}
\end{align}
 where $j=0,1,\cdots,M-1$, and ${\cal{N}}_{+,\pm1}^{(j+\alpha_1,\alpha_2)}$ are normalization constants.
 
It should be noticed that all of the ${\mathbb{Z}}_2$ eigenstates \eqref{Z2es} are not linearly independent. From Eq.\eqref{toruses} and properties of the theta function, the zero modes $\psi^{(j+\alpha_1,\alpha_2)}_{T^2+,0}$ satisfy
\begin{align}
\psi_{T^2+,0}^{(j+\alpha_1,\alpha_2)}(-z)&=\psi_{T^2+,0}^{(M-(j+\alpha_1),-\alpha_2)}(z) \notag \\
&=e^{-4\pi i \tfrac{j+\alpha_1}{M} \alpha_2}\psi_{T^2+,0}^{(M-(j+\alpha_1),\alpha_2)}(z).
\label{Torusz2}
\end{align}
Therefore, we have
 \begin{align}
\psi_{T^2/{\mathbb{Z}}_2+,0}^{(j+\alpha_1,\alpha_2)}(z)_{\pm1}=
{\cal{N}}_{+,\pm1}^{(j+\alpha_1,\alpha_2)}
[ \psi^{(j+\alpha_1,\alpha_2)}_{T^2+,0}(z)\pm e^{-4\pi i \tfrac{j+\alpha_1}{M} \alpha_2} \psi^{(M-(j+\alpha_1),\alpha_2)}_{T^2+,0}(z)].
\end{align}
It follows that the linearly independent ${\mathbb{Z}}_2$ eigenstates depend on $M$, $(\alpha_1,\alpha_2)$ and $\eta$, and are explicitly shown in Table \ref{T1}. The normalization constants ${\cal{N}}_{+,\eta}^{(j+\alpha_1,\alpha_2)}$ of the ${\mathbb{Z}}_2$ zero mode eigenstates are fixed by the orthonormality condition
\begin{align}
\int_{T^2/{\mathbb{Z}}_2}dz d\bar{z} \,\,[\psi_{T^2/{\mathbb{Z}}_2+,0}^{(j+\alpha_1,\alpha_2)}(z)_{\eta}]^{\ast}\psi_{T^2/{\mathbb{Z}}_2+,0}^{(k+\alpha_1,\alpha_2)}(z)_{\eta}
=\delta{^{jk}},
\label{Z2oc}
\end{align}
and turn out to depend on $j$ as well as $M$, $(\alpha_1,\alpha_2)$ and $\eta$, as shown in Table \ref{T2}. As we will see in the next section, the set of the independent ${\mathbb{Z}}_2$ eigenstates and the values of the normalization constants ${\cal{N}}_{+,\eta}^{(j+\alpha_1,\alpha_2)}$ are important to derive the index theorem on $T^2/{\mathbb{Z}}_2$.

 \begin{table}[h]
\centering
\begin{tabular}{c|c|c|c}
\hline
$(\alpha_1,\alpha_2)$&$M$&$\psi_{T^2/{\mathbb{Z}}_2,+,0}^{(j+\alpha_1,\alpha_2)}(z)_{+1}$&$\psi_{T^2/{\mathbb{Z}}_2,+,0}^{(j+\alpha_1,\alpha_2)}(z)_{-1}$\\ \hline
$(0,0)$&even&$\frac{M}{2}+1\quad(j=0,1,\cdots,\tfrac{M}{2})$&$\frac{M}{2}-1\quad(j=1,2,\cdots,\tfrac{M}{2}-1)$\\
$(0,0)$&odd&$\frac{M+1}{2}\quad(j=0,1,\cdots,\tfrac{M-1}{2})$&$\frac{M-1}{2}\quad(j=1,2,\cdots,\tfrac{M-1}{2})$\\ \hline
$(\tfrac{1}{2},0)$&even&$\frac{M}{2}\quad(j=0,1,\cdots,\tfrac{M}{2}-1)$&$\frac{M}{2}\quad(j=0,1,\cdots,\tfrac{M}{2}-1)$\\
$(\tfrac{1}{2},0)$&odd&$\frac{M+1}{2}\quad(j=0,1,\cdots,\tfrac{M-1}{2})$&$\frac{M-1}{2}\quad(j=0,1,\cdots,\tfrac{M-3}{2})$\\ \hline
$(0,\tfrac{1}{2})$&even&$\frac{M}{2}\quad(j=0,1,\cdots,\tfrac{M}{2}-1)$&$\frac{M}{2}\quad(j=1,2,\cdots,\tfrac{M}{2})$\\
$(0,\tfrac{1}{2})$&odd&$\frac{M+1}{2}\quad(j=0,1,\cdots,\tfrac{M-1}{2})$&$\frac{M-1}{2}\quad(j=1,2,\cdots,\tfrac{M-1}{2})$\\ \hline
$(\tfrac{1}{2},\tfrac{1}{2})$&even&$\frac{M}{2}\quad(j=0,1,\cdots,\tfrac{M}{2}-1)$&$\frac{M}{2}\quad(j=0,1,\cdots,\tfrac{M}{2}-1)$\\
$(\tfrac{1}{2},\tfrac{1}{2})$&odd&$\frac{M-1}{2}\quad(j=0,1,\cdots,\tfrac{M-3}{2})$&$\frac{M+1}{2}\quad(j=0,1,\cdots,\tfrac{M-1}{2})$\\ 
\hline
\end{tabular}
\caption{The number of zero modes on the $T^2/{\mathbb{Z}}_2$ orbifold.}
\label{T1}
\end{table}
 
\begin{table}[h]
\centering
{\tabcolsep = 4mm
\renewcommand{\arraystretch}{1.2}
\scalebox{0.80}{
\begin{tabular}{c|c|c|c}
\hline
$(\alpha_1,\alpha_2)$&$M$&${\cal{N}}_{+,+1}^{(j)}$&${\cal{N}}_{+,-1}^{(j)}$\\ \hline
$(0,0)$&even&$\frac{1}{\sqrt{2}} \,(j=0,\tfrac{M}{2}),\quad 1 \,(j={\rm{others}})$&$1 \,(j=1,2,\cdots,\tfrac{M}{2}-1)$\\
$(0,0)$&odd&$\frac{1}{\sqrt{2}} \,(j=0),\quad 1 \,(j={\rm{others}})$&$1 \,(j=1,2,\cdots,\tfrac{M-1}{2})$\\ \hline
$(\tfrac{1}{2},0)$&even&$1 \,(j=0,1,\cdots,\tfrac{M}{2}-1)$&$1 \,(j=0,1,\cdots,\tfrac{M}{2}-1)$\\
$(\tfrac{1}{2},0)$&odd&$\frac{1}{\sqrt{2}} \,(j=\tfrac{M-1}{2}),\quad 1 \,(j={\rm{others}})$&$1 \,(j=0,1,\cdots,\tfrac{M-3}{2})$\\ \hline
$(0,\tfrac{1}{2})$&even&$\frac{1}{\sqrt{2}} \,(j=0),\quad 1 \,(j={\rm{others}})$&$\frac{1}{\sqrt{2}} \,(j=\tfrac{M}{2}),\quad 1 \,(j={\rm{others}})$\\
$(0,\tfrac{1}{2})$&odd&$\frac{1}{\sqrt{2}} \,(j=0),\quad 1 \,(j={\rm{others}})$&$1 \,(j=1,2,\cdots,\tfrac{M-1}{2})$\\ \hline
$(\tfrac{1}{2},\tfrac{1}{2})$&even&$1 \,(j=0,1,\cdots,\tfrac{M}{2}-1)$&$1 \,(j=0,1,\cdots,\tfrac{M}{2}-1)$\\
$(\tfrac{1}{2},\tfrac{1}{2})$&odd&$1 \,(j=0,1,\cdots,\tfrac{M-3}{2})$&$\frac{1}{\sqrt{2}} \,(j=\tfrac{M-1}{2}),\quad 1 \,(j={\rm{others}})$\\ 
\hline
\end{tabular}
}
}
\caption{The normalization constants of zero modes on the $T^2/\mathbb{Z}_2$ orbifold.}
\label{T2}
\end{table} 
 
\subsection{$\mathbb{Z}_2$ eigen mode functions and mass spectrum}
The $\mathbb{Z}_2$ eigen mode functions $\psi_{T^2/\mathbb{Z}_2\pm,n}^{(j+\alpha_1,\alpha_2)}(z)_{\eta}$ satisfy
 \begin{align}
\begin{pmatrix}
-4D_z D_{\bar{z}} & 0\\
0 &-4 D_{\bar{z}}D_z 
\end{pmatrix}
\begin{pmatrix}
\psi_{T^2/\mathbb{Z}_2+,n}^{(j+\alpha_1,\alpha_2)}(z)_{\eta}\\
\psi_{T^2/\mathbb{Z}_2-,n}^{(j+\alpha_1,\alpha_2)}(z)_{\eta}
\end{pmatrix}
=m_n^2
\begin{pmatrix}
\psi_{T^2/\mathbb{Z}_2+,n}^{(j+\alpha_1,\alpha_2)}(z)_{\eta}\\
\psi_{T^2/\mathbb{Z}_2-,n}^{(j+\alpha_1,\alpha_2)}(z)_{\eta}
\end{pmatrix}
\label{Z2ev}
\end{align}
with
\begin{align}
\psi_{T^2/\mathbb{Z}_2+,n}^{(j+\alpha_1,\alpha_2)}(-z)_{\eta}
&=\eta\psi_{T^2/\mathbb{Z}_2+,n}^{(j+\alpha_1,\alpha_2)}(z)_{\eta}\,, \\
\psi_{T^2/\mathbb{Z}_2-,n}^{(j+\alpha_1,\alpha_2)}(-z)_{\eta}
&=-\eta\psi_{T^2/\mathbb{Z}_2-,n}^{(j+\alpha_1,\alpha_2)}(z)_{\eta}\,.
\end{align}
We note that if the $\mathbb{Z}_2$ eigenvalue of $\psi_{T^2/\mathbb{Z}_2+,n}^{(j+\alpha_1,\alpha_2)}(z){_\eta}$ is $\eta$, then that of $\psi_{T^2/\mathbb{Z}_2-,n}^{(j+\alpha_1,\alpha_2)}(z)_{\eta}$ has to be $\omega\eta=-\eta$. The additional factor $\omega=-1$ comes from a rotation matrix acting on 2d spinors, and also it is understood from the relation \eqref{MD2} with the property $\hat{a}_{+}(-z)=-\hat{a}_{+}(z)$.

\begin{figure}[!t]
\centering
\includegraphics[width=0.8\textwidth]{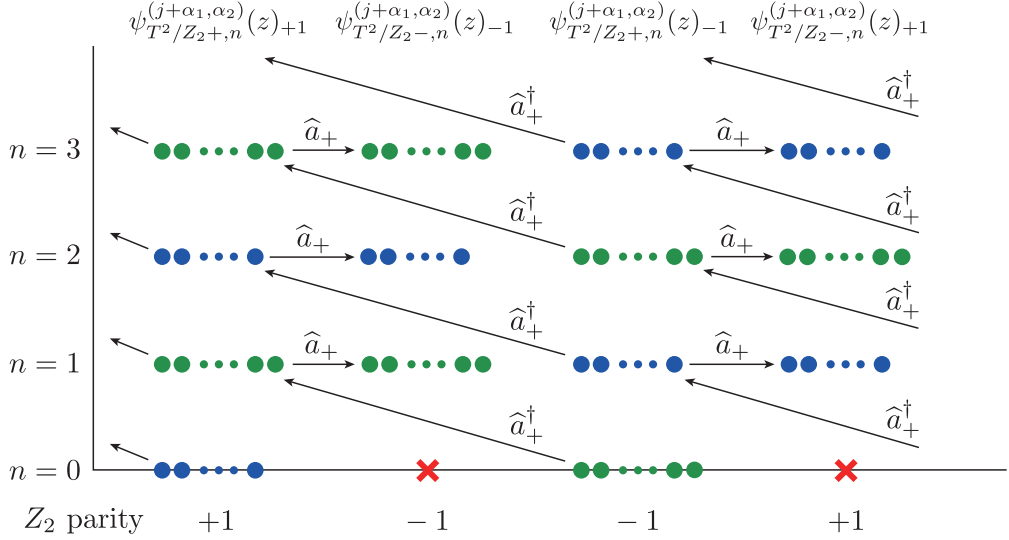}
\caption{The mass spectrum of $\psi_{T^2/{\mathbb{Z}}_2\pm,n}^{(j+\alpha_1,\alpha_2)}(z)_{\eta}$ for $M>0$. The blue (green)-filled circles correspond to zero modes with ${\mathbb{Z}}_2$ parity $+1$ ($-1$) and their Kaluza-Klein modes. The arrows mean that $\hat{a}_{+}^{\dagger}$ operates on the $n$th modes $\psi_{T^2/{\mathbb{Z}}_2+,n}^{(j+\alpha_1,\alpha_2)}(z)_{\eta}$ and creates the $(n+1)$th modes the next modes $\psi_{T^2/{\mathbb{Z}}_2+,n+1}^{(j+\alpha_1,\alpha_2)}(z)_{-\eta}$ with {the} opposite $\mathbb{Z}_N$ eigenvalue, and also that $\hat{a}_{+}$ operates on the $n$th modes $\psi_{T^2/{\mathbb{Z}}_2+,n}^{(j+\alpha_1,\alpha_2)}(z)_{\eta}$ and creates the $n$th modes $\psi_{T^2/{\mathbb{Z}}_2-,n}^{(j+\alpha_1,\alpha_2)}(z)_{-\eta}$.}
\label{figureZ2}
\end{figure}

In terms of the zero modes $\psi_{T^2/\mathbb{Z}_2+,0}^{(j+\alpha_1,\alpha_2)}(z)_{\eta}$ on the $T^2/\mathbb{Z}_2$ orbifold, the $\mathbb{Z}_2$ massive mode functions $\psi_{T^2/\mathbb{Z}_2\pm,n}^{(j+\alpha_1,\alpha_2)}(z)_{\eta}\,\,(n=1,2,\cdots)$ can be constructed as
\begin{align}
\psi_{T^2/{\mathbb{Z}}_2+,n}^{(j+\alpha_1,\alpha_2)}(z)_{\eta}&\equiv  \frac{1}{\sqrt{n!}}[\hat{a}_{+}^{\dagger}(z)]^n\psi_{T^2/{\mathbb{Z}}_2+,0}^{(j+\alpha_1,\alpha_2)}(z)_{\omega^n \eta} \notag \\
 &= {\cal{N}}_{+,\omega^n \eta}^{(j+\alpha_1,\alpha_2)}
 \sum_{l=0}^{1}{\eta}^{-l} \psi^{(j+\alpha_1,\alpha_2)}_{T^2+,n}(\omega^l z),\label{Z2KKmode1}\\
\psi_{T^2/{\mathbb{Z}}_2-,n}^{(j+\alpha_1,\alpha_2)}(z)_{\eta}&=\frac{2}{m_n}D_{\bar{z}}\psi_{T^2/{\mathbb{Z}}_2+,n}^{(j+\alpha_1,\alpha_2)}(z)_{\eta}  \notag \\
&=- i {\cal{N}}_{+,\omega^n \eta}^{(j+\alpha_1,\alpha_2)}
 \sum_{l=0}^{1}{(\omega \eta)}^{-l} \psi^{(j+\alpha_1,\alpha_2)}_{T^2+,n-1}(\omega^l z),
\label{Z2KKmode}
\end{align}
 where $\omega=-1$ and $\eta=\pm1$. We notice that the number of the degeneracy of $\psi_{T^2/{\mathbb{Z}}_2\pm,n}^{(j+\alpha_1,\alpha_2)}(z)_{\eta}$ with a fixed $\eta$ depends on $n$, in general, because that of $\psi_{T^2/{\mathbb{Z}}_2+,0}^{(j+\alpha_1,\alpha_2)}(z)_{\omega^n \eta}$ does from Table \ref{T1}. The mass spectrum of $\psi_{T^2/{\mathbb{Z}}_2\pm,n}^{(j+\alpha_1,\alpha_2)}(z)_{\eta}$ is shown in Figure \ref{figureZ2}.
 
\section{Index theorem on $T^2/{\mathbb{Z}}_2$ orbifold with magnetic flux}
In this section, we derive the index on the $T^2/{\mathbb{Z}}_2$ orbifold with magnetic flux by using the trace formula
\begin{align}
{\rm{Ind}}(i\slashed{D})_{\eta}=\lim_{\rho \to \infty} {\rm{tr}}[\sigma_3 e^{\slashed{D}^2/\rho^2}]_{\eta},
\label{Ind}
\end{align}
where the subscript $\eta$ means that the trace on the right-hand side of Eq.\eqref{Ind} should be restricted in the functional space spanned by the mode functions $\{\psi_{T^2/{\mathbb{Z}}_2\pm,n}^{(j+\alpha_1,\alpha_2)}(z)_{\eta}\}$. The regularization factor $e^{\slashed{D}^2/\rho^2}$ is introduced in the trace of Eq.\eqref{Ind} to make the trace of $\sigma_3$ well defined. Using the relation 
$\slashed{D}^2{=\sigma^a \sigma^b D_a D_b}=D^2{-}\tfrac{iq}{4}[\sigma^{a},\sigma^{b}]F_{ab}$, where $D^2=4D_{z}D_{\bar{z}}-2\pi M/{{\rm{Im}} \tau}$ and $\sigma^{a}(a=1,2)$ {is} {the} Pauli matrix, we expand Eq.\eqref{Ind} around $D^2$ as follows:
\begin{align}
{\rm{Ind}}(i\slashed{D})_{\eta}=\lim_{\rho \to \infty} \left\{{\rm{tr}}[\sigma_3 e^{{D}^2/\rho^2}]+\tfrac{1}{\rho^2} {\rm{tr}}[\sigma_3 ({-}\tfrac{iq}{4})[\sigma^{a},\sigma^{b}]F_{ab}e^{{D}^2/\rho^2}]+{\cal{O}}(\rho^{-4})\right\}.
\label{Ind2}
\end{align}

The second term of Eq.\eqref{Ind2} can be evaluated by the Fujikawa method~\cite{PhysRevLett.42.1195,PhysRevD.22.1499} and is given by $M/2$, where the factor $1/2$ comes from the fact that the area of the $T^2/{\mathbb{Z}}_2$ orbifold is $1/2$ of that of the torus $T^2$. The third term of ${\cal{O}}(\rho^{-4})$ will vanish in the limit of $\rho\to \infty$.

The first term of Eq.\eqref{Ind2} is usually disregarded because ${\rm{tr}}[\sigma_3]$ is expected to vanish with $\rho\to \infty$. This is not, however, the case since the $T^2/{\mathbb{Z}}_2$ orbifold has singular points, which correspond to the fixed points, i.e. $z=0,1/2,\tau/2$ and $(1+\tau)/2$, so that we cannot directly apply the Atiyah-Singer index theorem  for the $T^2/{\mathbb{Z}}_2$ orbifold. One of our main purposes of this paper is to show that the first term of Eq.\eqref{Ind2} does not vanish and gives the desired result as the index theorem, as we will see below.

By using the complete set of the mode functions $\{\,\psi_{T^2/{\mathbb{Z}}_2+,n}^{(j+\alpha_1,\alpha_2)}(z)_{\eta}$ $\,\,(n=0,1,\cdots)\,\}$ and $\{\psi_{T^2/{\mathbb{Z}}_2-,n}^{(j+\alpha_1,\alpha_2)}(z)_{\eta}$ $(n=1,2,\cdots)\}$, the first term of Eq.\eqref{Ind2} can be expressed as
\begin{align}
&\lim_{\rho \to \infty} {\rm{tr}}[\sigma_3 e^{{D}^2/\rho^2}]_{\eta} \notag \\
&=
\lim_{\rho \to \infty}\int_{T^2/{\mathbb{Z}}_2}dz d{\bar{z}} \lim_{z^{\prime} \to z}
e^{{D}^2(z)/\rho^2}\left\{\sum_{n=0}^{\infty} \sum_{j}
\psi_{T^2/{\mathbb{Z}}_2+,n}^{(j+\alpha_1,\alpha_2)}(z)_{\eta}[\psi_{T^2/{\mathbb{Z}}_2+,n}^{(j+\alpha_1,\alpha_2)}(z^{\prime})_{\eta}]^{\ast} \right.\notag \\
&\left.\hspace{150pt} -\sum_{n=1}^{\infty} \sum_{j}
\psi_{T^2/{\mathbb{Z}}_2-,n}^{(j+\alpha_1,\alpha_2)}(z)_{ \eta}[\psi_{T^2/{\mathbb{Z}}_2-,n}^{(j+\alpha_1,\alpha_2)}(z^{\prime})_{\eta}]^{\ast}\right\},
\label{Indz2}
\end{align}
where $j$ denotes the label of the degeneracy of the linearly independent ${\mathbb{Z}}_2$ eigen functions, which are given in Table \ref{T1}. It is interesting to point out that each of the first and the second terms in Eq.\eqref{Indz2} is divergent, but their combination in Eq.\eqref{Indz2} is finite. Indeed, Eq.\eqref{Indz2} can be evaluated as
\begin{align}
\lim_{\rho \to \infty} {\rm{tr}}[\sigma_3 e^{{D}^2/\rho^2}]_{\eta}
=\frac{1}{2}\int_{T^2} dz d{\bar{z}} \sum_{n=0}^{\infty} \sum_{j=0}^{M-1} \eta (1-\omega)
 \psi^{(j+\alpha_1,\alpha_2)}_{T^2+,n}(z)[ \psi^{(j+\alpha_1,\alpha_2)}_{T^2+,n}(\omega z)]^{\ast}.
\label{Ind3}
\end{align}
We note that there appear the wavefunctions defined on $T^2$ (but not on $T^2/{\mathbb{Z}}_2$) as well as the integration over the area of $T^2$ in Eq.\eqref{Ind3}.

The proof of Eq.\eqref{Ind3} is lengthy because we need to verify the expression \eqref{Ind3} for every case of $M=$ even/odd, $ (\alpha_1,\alpha_2)=(0,0),(1/2,0),(0,1/2),(1/2,1/2)$ and $\eta=\pm 1$, separately. {In Appendix B}, we show Eq.\eqref{Ind3} only for the case of $(\alpha_1,\alpha_2)=(0,0)$ with {$M=\rm{even/odd}$}  and $\eta=\pm 1$. Other cases can be verified in a similar manner.

Since $\{\psi_{T^2+,n}^{(j+\alpha_1,\alpha_2)}(z)\,\,(n=0,1,\cdots;\,\,j=0,1,\cdots,M-1)\}$ forms the complete set of the wavefunctions on the torus $T^2$, Eq.\eqref{Ind3} can be expressed as
\begin{align}
&\lim_{\rho \to \infty} {\rm{tr}}[\sigma_3 e^{{D}^2/\rho^2}]_{\eta} 
=\frac{1}{2}\int_{T^2} dz d{\bar{z}} \,\eta (1-\omega)
\delta^2_{T^2}(z,\omega z),
\label{Inddelta}
\end{align}
where
\begin{align}
\delta_{T^2}^2(z,w)\equiv\sum_{n=0}^{\infty} \sum_{j=0}^{M-1}\psi_{T^2+,n}^{(j+\alpha_1,\alpha_2)}(z)
[\psi_{T^2+,n}^{(j+\alpha_1,\alpha_2)}(w)]^{\ast}.
\label{CO}
\end{align}
To proceed further, we need the explicit representation of the {specific} delta function $\delta_{T^2}^2(z,w)$ on the torus $T^2$. The {specific} delta function $\delta_{T^2}^2(z,w)$ defined by Eq.\eqref{CO} should satisfy the relations
\begin{align}
&\delta_{T^2}^2(z+1,w)=e^{iq\Lambda_1(z)+i2\pi \alpha_1}\delta_{T^2}^2(z,w),\label{delta1} \\
&\delta_{T^2}^2(z+\tau,w)=e^{iq\Lambda_2(z)+i2\pi \alpha_2}\delta_{T^2}^2(z,w),\\
&\delta_{T^2}^2(z,w+1)=\delta_{T^2}^2(z,w)e^{-iq\Lambda_1(w)-i2\pi \alpha_1},\\
&\delta_{T^2}^2(z,w+\tau)=\delta_{T^2}^2(z,w)e^{-iq\Lambda_2(w)-i2\pi \alpha_2},\\
&\int_{T^2} dw d{\bar{w}} \,\delta_{T^2}^2(z,w) \psi_{T^2+,n}^{(j+\alpha_1,\alpha_2)}(w)=\psi_{T^2+,n}^{(j+\alpha_1,\alpha_2)}(z), \\
&\int_{T^2} dz d{\bar{z}}\, [ \psi_{T^2+,n}^{(j+\alpha_1,\alpha_2)}(z)]^{\ast}\delta_{T^2}^2(z,w)=[\psi_{T^2+,n}^{(j+\alpha_1,\alpha_2)}(w)]^{\ast}.
\label{delta6}
\end{align}
It turns out that the {specific} delta function $\delta_{T^2}^2(z,w)$ satisfying Eqs.\eqref{delta1}$-$\eqref{delta6} can be represented {in terms of the standard delta function $\delta^2(z)$} as
\begin{align}
\delta_{T^2}^2(z,w)
=
\sum_{m,n\in {\mathbb{Z}}}e^{i\theta_{m,n}(z)}\delta^2(z-w-m-n\tau),
\label{delta}
\end{align}
where
\begin{align}
\theta_{m,n}(z)=mq\Lambda_1(z)+2\pi m\alpha_1+nq\Lambda_2(z)+2\pi n\alpha_2+\pi M mn.
\label{thetamn}
\end{align}
The non-trivial phase $\theta_{m,n}(z)$ is crucially important to derive the correct index and is found to be directly related to winding numbers at fixed points on the orbifolds, as we will see in the Appendix {C}.

Using the expression \eqref{delta} of the delta function with the nontrivial phase \eqref{thetamn}, we can rewrite Eq.\eqref{Inddelta} as
\begin{align}
\lim_{\rho \to \infty} {\rm{tr}}[\sigma_3 e^{{D}^2/\rho^2}]_{\eta} 
&=\int_{T^2} dz d{\bar{z}} \sum_{m,n \in {\mathbb{Z}}}\eta e^{i\theta_{m,n}(z)}\delta^2(z-\omega z-m-n\tau) \notag \\
&=\frac{1}{4}\int_{T^2} dz d{\bar{z}} \sum_{m,n \in {\mathbb{Z}}}\eta e^{i\theta_{m,n}(z)}\delta^2(z-(m+n\tau)/2),
\label{Inddelta2}
\end{align}
where we have used $\omega=-1$.

By taking the fundamental domain of the torus $T^2$ to be $-\varepsilon \leq y_1,y_2<1-\varepsilon$ with a small positive number {$\varepsilon$} instead of $0\leq y_1,y_2<1$, the values of $(m,n)$, which remain in the summation of Eq.\eqref{Inddelta2} after the $z$-integration, are given by
\begin{align}
(m,n)=(0,0),(1,0),(0,1),(1,1).
\label{z2mn}
\end{align}
Then, we have
\begin{align}
\lim_{\rho \to \infty} {\rm{tr}}[\sigma_3 e^{{D}^2/\rho^2}]_{\eta} 
= \frac{1}{{2 \times 2}}\int_{T^2}dz d{\bar{z}} \sum_{p=1}^4 W_p \delta^2(z-z_p^f),
\label{IndWp}
\end{align}
where $z_p^f\,\,(p=1,2,3,4)$ are given by
\begin{align}
z_1^f=0,\quad z_2^f=1/2,\quad  z_3^f=\tau/2,\quad  z_4^f=(1+\tau)/2,
\end{align}
which are nothing but the fixed points on the $T^2/{\mathbb{Z}}_2$ orbifold. The coefficients $W_p$ in Eq.\eqref{IndWp} are
\begin{align}
W_1&=\eta e^{i\theta_{m,n}(z)}\Big{|}^{(m,n)=(0,0)}_{z=z^f_1=0}, \\
W_2&=\eta e^{i\theta_{m,n}(z)}\Big{|}^{(m,n)=(1,0)}_{z=z^f_2=1/2}, \\
W_3&=\eta e^{i\theta_{m,n}(z)}\Big{|}^{(m,n)=(0,1)}_{z=z^f_3=\tau/2}, \\
W_4&=\eta e^{i\theta_{m,n}(z)}\Big{|}^{(m,n)=(1,1)}_{z=z^f_4=(1+\tau)/2}.
\label{Z2coe}
\end{align}
The explicit values of $W_p\,(p=1,2,3,4)$ are summarized in Table \ref{T3}.
\begin{table}[!t]
\centering
{\tabcolsep = 4mm
\renewcommand{\arraystretch}{1.2}
\scalebox{0.80}{
\begin{tabular}{ccc|c:c:c:c|c|c} \hline
flux & parity & twist & \multicolumn{4}{c|}{coefficients of the delta functions}  &sum of coefficients & index\\
$M$ & $\eta$ & $(\alpha_1, \alpha_2)$ & $~W_{1}~$ & $~W_{2}~$ & $~W_{3}~$ & $W_{4}$ &  $\sum_{p=1}^4 W_p$ &$M/2+\sum_{p=1}^4 W_p/4$\\ \hline
$2m+1$ & $+1$ & $(0,0)$ & $+1$ & $+1$ & $+1$ & $-1$& $+2$&$(M+1)/2$\\
&& $(\tfrac12, 0)$ & $+1$ & $-1$ & $+1$ & $+1$& $+2$& $(M+1)/2$ \\
&& $(0, \tfrac12)$ & $+1$ & $+1$ & $-1$ & $+1$& $+2$& $(M+1)/2$ \\
&& $(\tfrac12, \tfrac12)$ & $+1$ & $-1$ & $-1$ & $-1$& $-2$  & $(M-1)/2$ \\ \cline{2-9}
& $-1$ & $(0,0)$ & $-1$ & $-1$ & $-1$ & $+1$& $-2$& $(M-1)/2$ \\
&& $(\tfrac12,0)$ & $-1$ & $+1$ & $-1$ & $-1$& $-2$ & $(M-1)/2$ \\
&& $(0, \tfrac12)$ & $-1$ & $-1$ & $+1$ & $-1$& $-2$& $(M-1)/2$ \\
&& $(\tfrac12, \tfrac12)$ & $-1$ & $+1$ & $+1$ & $+1$& $+2$& $(M+1)/2$ \\ \hline
$2m+2$ & $+1$ & $(0,0)$ & $+1$ & $+1$ & $+1$ & $+1$& $+4$ & $M/2+1$ \\
&& $(\tfrac12, 0)$ & $+1$ & $-1$ & $+1$ & $-1$& $0$& $M/2$ \\
&& $(0, \tfrac12)$ & $+1$ & $+1$ & $-1$ & $-1$& $0$& $M/2$ \\
&& $(\tfrac12, \tfrac12)$ & $+1$ & $-1$ & $-1$ & $+1$& $0$& $M/2$ \\ \cline{2-9}
& $-1$ & $(0,0)$ & $-1$ & $-1$ & $-1$ & $-1$& $-4$& $M/2-1$ \\
&& $(\tfrac12, 0)$& $-1$ & $+1$ & $-1$ & $+1$& $0$& $M/2$ \\
&& $(0, \tfrac12)$ & $-1$ & $-1$ & $+1$ & $+1$& $0$& $M/2$ \\
&& $(\tfrac12, \tfrac12)$ & $-1$ & $+1$ & $+1$ & $-1$& $0$ & $M/2$ \\ \hline
\end{tabular}
}
}
\caption{The coefficients {$W_p$} of {the} delta functions in {Eq.}\eqref{IndWp} and {the} index {for the $T^2/{\mathbb{Z}}_2$ orbifold}.
}
\label{T3}
\end{table}  

We have succeeded in deriving the index of the $T^2/{\mathbb{Z}}_2$ orbifold, and the result\red{s} shown in Table \ref{T3} are found to be consistent with the number of the zero modes in Table \ref{T1}. Any physical meaning of the coefficient $W_p$ is not, however, clear at this moment. In Section 6, we will clarify that $W_p$ is related to the winding numbers at the fixed point {$z_p^f$} on the $T^2/{\mathbb{Z}}_2$ orbifold.

\section{Index theorem on $T^2/{\mathbb{Z}}_N\,\,(N=3,4,6)$ orbifolds with magnetic flux}
The $T^2/{\mathbb{Z}}_N\,\,(N=3,4,6)$ orbifolds are defined by the identification
\begin{align}
z \,\, \sim \,\,\omega z \qquad (\omega=e^{i2\pi/N})
\end{align}
in addition to the torus identification \eqref{torusid}. The SS phase $(\alpha_1,\alpha_2)$ has to be quantized as 
\begin{gather}
\alpha = \alpha_{1} = \alpha_{2} = 
\begin{cases} 
0, 1/3, 2/3 & \quad (M = \textrm{even})\\
1/6, 3/6, 5/6 & \quad (M = \textrm{odd})
\end{cases} \qquad  {\rm{for}}\,\, T^2/\mathbb{Z}_3,
\label{Z3_SSphase} \\
\alpha = \alpha_{1} = \alpha_{2}
= 0, 1/2 \qquad  {\rm{for}}\,\, T^{2}/\mathbb{Z}_{4},
\label{Z4_SSphase} \\
\alpha = \alpha_{1} = \alpha_{2} = 
\begin{cases} 
0 & \quad (M = \textrm{even})\\
1/2 & \quad (M = \textrm{odd})
\end{cases} \qquad  {\rm{for}}\,\, T^2/\mathbb{Z}_6.
\label{Z6_SSphase}
\end{gather}

The $\mathbb{Z}_N$ eigenstates $\psi_{T^2/\mathbb{Z}_N\pm,n}^{(j+\alpha_1,\alpha_2)}(z)_{\eta}$ satisfy the eigenvalue equations
\begin{align}
\psi_{T^2/\mathbb{Z}_N+,n}^{(j+\alpha_1,\alpha_2)}(\omega z)_{\eta}&=\eta
\psi_{T^2/\mathbb{Z}_N+,n}^{(j+\alpha_1,\alpha_2)}(z)_{\eta},\label{ZNes1}\\
\psi_{T^2/\mathbb{Z}_N-,n}^{(j+\alpha_1,\alpha_2)}(\omega z)_{\eta}&=\omega \eta
\psi_{T^2/\mathbb{Z}_N-,n}^{(j+\alpha_1,\alpha_2)}(z)_{\eta},
\label{ZNes}
\end{align}
where $\eta=\omega^l\,(l=0,1,\cdots,N-1)$ denotes the ${\mathbb{Z}}_N$ eigenvalue with $\omega=e^{i2\pi/N}$. In terms of the zero mode functions $\psi_{T^2+,0}^{(j+\alpha_1,\alpha_2)}(z)$ on the torus $T^2$, the ${\mathbb{Z}}_N$ zero mode functions $\psi_{T^2/\mathbb{Z}_N+,0}^{(j+\alpha_1,\alpha_2)}(z)_{\eta}$ on the $T^2/{\mathbb{Z}}_N$ orbifolds {can} be constructed as
\begin{align}
 \psi_{T^2/{\mathbb{Z}}_N+,0}^{(j+\alpha_1,\alpha_2)}( z)_{\eta}\equiv
{\cal{N}}_{+,\eta}^{(j+\alpha_1,\alpha_2)}\sum_{l=0}^{N-1}
{\eta}^{-l} \psi^{(j+\alpha_1,\alpha_2)}_{T^2+,0}(\omega^l z).
\label{ZNtransf}
\end{align}

The problem is here that all of $\psi_{T^2/{\mathbb{Z}}_N+,0}^{(j+\alpha_1,\alpha_2)}(z)_{\eta}$ for $j=0,1,\cdots,M-1$ with a fixed $\eta$ are not always linearly independent. Although we have obtained the complete set of the linearly independent ${\mathbb{Z}}_2$ eigenfunctions on the $T^2/{\mathbb{Z}}_2$ orbifold in Section 3, it is highly nontrivial to construct complete sets of ${\mathbb{Z}}_N$ eigenfunctions on the $T^2/{\mathbb{Z}}_N\,\,(N=3,4,6)$ orbifolds except for some small $M$ \cite{Abe:2013bca}. Indeed, the complete sets of  ${\mathbb{Z}}_N$ eigenfunctions on the $T^2/{\mathbb{Z}}_N\,\,(N=3,4,6)$ orbifolds for general $M$ are unknown, so that we cannot follow the analysis given in Sections 3 and 4 to derive the index theorem for $T^2/{\mathbb{Z}}_N\,\,(N=3,4,6)$.

To evaluate $\lim_{\rho\to\infty}{\rm{tr}}[\sigma_3 e^{D^2/\rho^2}]_{\eta}$ for the $T^2/{\mathbb{Z}}_N\,\,(N=3,4,6)$ orbifolds, we extend the relation \eqref{Ind3} for $T^2/{\mathbb{Z}}_2$ to
 \begin{align}
\lim_{\rho \to \infty} {\rm{tr}}[\sigma_3 e^{{D}^2/\rho^2}]_{\eta}
=\frac{1}{N}\int_{T^2} d z d{\bar{z}} \sum_{n=0}^{\infty} \sum_{j=0}^{M-1} \sum_{l=1}^{N-1} \eta^l (1-\omega^l)
 \psi^{(j+\alpha_1,\alpha_2)}_{T^2+,n}(z)[ \psi^{(j+\alpha_1,\alpha_2)}_{T^2+,n}(\omega^l z)]^{\ast}.
\label{Ind57}
\end{align}
This formula is true for the case of $M=0$, as shown in~\cite{Sakamoto:2020vdy}, and we can also verify Eq.\eqref{Ind57} for some small $M$ by explicitly constructing complete sets of ${\mathbb{Z}}_N$ eigenfunctions on the $T^2/{\mathbb{Z}}_N\,\,(N=3,4,6)$ orbifolds. Unfortunately, we have not succeeded in proving the relation \eqref{Ind57} for arbitrary $M$. We leave the proof of Eq.\eqref{Ind57} for future work. In the following, we will show that the relation \eqref{Ind57} leads to the correct results.

By use of the completeness relation \eqref{CO} with the representation \eqref{delta}, Eq.\eqref{Ind57} becomes
\begin{align}
\lim_{\rho \to \infty} {\rm{tr}}[\sigma_3 e^{{D}^2/\rho^2}]_{\eta} 
=\frac{1}{N}\int_{T^2} d z d{\bar{z}}  \sum_{l=1}^{N-1} \eta^l (1-\omega^l)
\sum_{m,n\in{\mathbb{Z}}} 
e^{i\theta_{m,n}(z)}\delta^2(z-\omega^l z-m-n\tau).
\label{Ind58}
\end{align}
We evaluate Eq.\eqref{Ind58} for $N=3,4$ and 6, separately.

\subsection{$T^2/{\mathbb{Z}}_3$}
In the case of $N=3$, we obtain 
\begin{align}
&\lim_{\rho \to \infty} {\rm{tr}}[\sigma_3 e^{{D}^2/\rho^2}]_{\eta} \notag \\
&=\frac{1}{3}\int_{T^2} d z d{\bar{z}}  \sum_{k=1}^{2} \sum_{m_k,n_k\in{\mathbb{Z}}} \eta^k(1-\omega^k)
e^{i\theta_{m_k,n_k}(z)}\delta^2(z-\omega^k z-m_k-n_k\tau).
\label{Ind59}
\end{align}
Since  we have taken the fundamental domain of the torus $T^2$ to be $-\varepsilon \leq y_1,y_2<1-\varepsilon$, the values of $(m_1,n_1)$ and $(m_2,n_2)$, which remain in the summations of Eq.\eqref{Ind59} after the $z$-integration, are given by
\begin{align}
(m_1,n_1)=(0,0),(1,0),(1,1),\\
(m_2,n_2)=(0,0),(1,1),(0,1).
\end{align}
Then, it follows that Eq.\eqref{Ind59} reduces to
\begin{align}
\lim_{\rho \to \infty} {\rm{tr}}[\sigma_3 e^{{D}^2/\rho^2}]_{\eta}
&=\frac{1}{2\times 3}\int_{T^2} dz d{\bar{z}} \sum_{p=1}^3 W_p \delta^2(z-{z_p^f}),
\label{Ind512}
\end{align}
where we have used the relations
\begin{align}
\delta^2(z-\omega z -m_1-n_1\tau)&=\frac{1}{3}\delta^2(z-(2m_1-n_1)/3-(m_1+n_1)\tau/3), \\
\delta^2(z-\omega^2 z -m_2-n_2\tau)&=\frac{1}{3}\delta^2(z-(m_2+n_2)/{3}-(-m_2+2n_2)\tau/{3}).
\end{align}
The ${z_p^f}\,\,(p=1,2,3)$ are the fixed points of the $T^2/{\mathbb{Z}}_3$ orbifold, i.e.
\begin{align}
z_1^f=0,\quad z_2^f=\frac{1}{3}(2+\tau),\quad z_3^f=\frac{1}{3}(1+2\tau)
\label{Z3fixed}
\end{align}
and the coefficients $W_p\,\,(p=1,2,3)$ are given by
\begin{align}
W_1&=\frac{2}{3}\bigl\{\eta(1-\omega)
e^{i\theta_{m_1,n_1}(z)}{\big{|}}^{(m_1,n_1)=(0,0)}_{z={z_1^f}}
+\eta^2(1-\omega^2)e^{i\theta_{m_2,n_2}(z)}{\big{|}}^{(m_2,n_2)=(0,0)}_{z={z_1^f}}\bigr\},
\label{IndW31}\\ 
W_2&=\frac{2}{3}\bigl\{\eta(1-\omega)e^{i\theta_{m_1,n_1}(z)}{\big{|}}^{(m_1,n_1)=(1,0)}_{z={z_2^f}}
+\eta^2(1-\omega^2)e^{i\theta_{m_2,n_2}(z)}{\big{|}}^{(m_2,n_2)=(1,1)}_{z={z_2^f}}\bigr\},\\
W_3&=\frac{2}{3}\bigl\{\eta(1-\omega)e^{i\theta_{m_1,n_1}(z)}{\big{|}}^{(m_1,n_1)=(1,1)}_{z={z_3^f}}
+\eta^2(1-\omega^2)e^{i\theta_{m_2,n_2}(z)}{\big{|}}^{(m_2,n_2)=(0,1)}_{z={z_3^f}}\bigr\}.
\label{IndW33}
\end{align}
The explicit values of $W_p\,\,(p=1,2,3)$ are summarized in Table \ref{T4} and \ref{T5}.
\begin{table}[!ht]
	\centering
	{\tabcolsep = 3mm
		\renewcommand{\arraystretch}{1.2}
		\scalebox{0.80}{
			\begin{tabular}{ccc|c:c:c|c|c} \hline
			 flux& parity& twist & \multicolumn{3}{c|}{coefficients of the delta functions } &sum of coefficients & index\\
				$M$ & $\eta$ & $\alpha$ & $\quad W_1 \quad$ & $\quad W_2 \quad$ & $\quad W_3 \quad$ & $\sum_{p=1}^3 W_p$ &
				$M/3+\sum_{p=1}^3 W_p/6$\\ \hline
				$6m+1$ & $1$ & $1/6$ & $+2$ & $0$ & $+2$ & $+4$ & $(M+2)/3$ \\
				&& $1/2$ & $+2$ & $-2$ & $-2$ & $-2$  & $(M-1)/3$ \\
				&& $5/6$ & $+2$ & $+2$ & $0$ & $+4$ & $(M+2)/3$  \\ \cline{2-8}
				& $\omega$ & $1/6$ & $0$ & $-2$ & $0$ & $-2$ & $(M-1)/3$  \\
				&& $1/2$ & $0$ & $+2$ & $+2$ & $+4$ & $(M+2)/3$  \\
				&& $5/6$ & $0$ & $0$ & $-2$ & $-2$ & $(M-1)/3$  \\ \cline{2-8}
				& $\omega^2$ & $1/6$ & $-2$ & $+2$ & $-2$ & $-2$ & $(M-1)/3$  \\
				&& $1/2$ & $-2$ & $0$ & $0$ & $-2$ & $(M-1)/3$  \\
				&& $5/6$ & $-2$ & $-2$ & $+2$ & $-2$ & $(M-1)/3$  \\ \hline
				$6m+2$ & $1$ & $0$ & $+2$ & $0$ & $0$ & $+2$ & $(M+1)/3$  \\
				&& $1/3$ & $+2$ & $-2$ & $+2$ & $+2$ & $(M+1)/3$  \\
				&& $2/3$ & $+2$ & $+2$ & $-2$ & $+2$ & $(M+1)/3$  \\ \cline{2-8}
				& $\omega$ & $0$ & $0$ & $-2$ & $-2$ & $-4$ & $(M-2)/3$  \\
				&& $1/3$ & $0$ & $+2$ & $0$ & $+2$ & $(M+1)/3$  \\
				&& $2/3$ & $0$ & $0$ & $+2$ & $+2$ & $(M+1)/3$  \\ \cline{2-8}
				& $\omega^2$ & $0$ & $-2$ & $+2$ & $+2$ & $+2$ & $(M+1)/3$  \\
				&& $1/3$ & $-2$ & $0$ & $-2$ & $-4$ & $(M-2)/3$  \\
				&& $2/3$ & $-2$ & $-2$ & $0$ & $-4$ & $(M-2)/3$  \\ \hline
				$6m+3$ & $1$ & $1/6$ & $+2$ & $-2$ & $0$ & $0$ & $M/3$  \\
				&& $1/2$ & $+2$ & $+2$ & $+2$ & $+6$ & $M/3+1$  \\
				&& $5/6$ & $+2$ & $0$ & $-2$ & $0$ & $M/3$  \\ \cline{2-8}
				& $\omega$ & $1/6$ & $0$ & $+2$ & $-2$ & $0$ & $M/3$  \\
				&& $1/2$ & $0$ & $0$ & $0$ & $0$ & $M/3$  \\
				&& $5/6$ & $0$ & $-2$ & $+2$ & $0$ & $M/3$  \\ \cline{2-8}
				& $\omega^2$ & $1/6$ & $-2$ & $0$ & $+2$ & $0$ & $M/3$  \\
				&& $1/2$ & $-2$ & $-2$ & $-2$ & $-6$ & $M/3-1$  \\
				&& $5/6$ & $-2$ & $+2$ & $0$ & $0$ & $M/3$  \\ \hline
			\end{tabular}
		}
	}
	\caption{The coefficients {$W_p$} of {the} delta functions in {Eqs.}\eqref{IndW31}-\eqref{IndW33} and {the} index {for the $T^2/{\mathbb{Z}}_3$ orbifold}.
		}
	\label{T4}
\end{table}
\begin{table}[!ht]
	\centering
	{\tabcolsep = 3mm
		\renewcommand{\arraystretch}{1.2}
		\scalebox{0.80}{
			\begin{tabular}{ccc|c:c:c|c|c} \hline
			 flux& parity& twist & \multicolumn{3}{c|}{coefficients of the delta functions } &sum of coefficients & index\\
				$M$ & $\eta$ & $\alpha$ & $\quad W_1 \quad$ & $\quad W_2 \quad$ & $\quad W_3 \quad$ & $\sum_{p=1}^3 W_p$ &
				$M/3+\sum_{p=1}^3 W_p/6$\\ \hline
				$6m+4$ & $1$ & $0$ & $+2$ & $-2$ & $-2$ & $-2$ & $(M-1)/3$ \\
				&& $1/3$ & $+2$ & $+2$ & $0$ & $+4$  & $(M+2)/3$ \\
				&& $2/3$ & $+2$ & $0$ & $+2$ & $+4$ & $(M+2)/3$  \\ \cline{2-8}
				& $\omega$ & $0$ & $0$ & $+2$ & $+2$ & $+4$ & $(M+2)/3$  \\
				&& $1/3$ & $0$ & $0$ & $-2$ & $-2$ & $(M-1)/3$  \\
				&& $2/3$ & $0$ & $-2$ & $0$ & $-2$ & $(M-1)/3$  \\ \cline{2-8}
				& $\omega^2$ & $0$ & $-2$ & $0$ & $0$ & $-2$ & $(M-1)/3$  \\
				&& $1/3$ & $-2$ & $-2$ & $+2$ & $-2$ & $(M-1)/3$  \\
				&& $2/3$ & $-2$ & $+2$ & $-2$ & $-2$ & $(M-1)/3$  \\ \hline
				$6m+5$ & $1$ & $1/6$ & $+2$ & $+2$ & $-2$ & $+2$ & $(M+1)/3$  \\
				&& $1/2$ & $+2$ & $0$ & $0$ & $+2$ & $(M+1)/3$  \\
				&& $5/6$ & $+2$ & $-2$ & $+2$ & $+2$ & $(M+1)/3$  \\ \cline{2-8}
				& $\omega$ & $1/6$ & $0$ & $0$ & $+2$ & $+2$ & $(M+1)/3$  \\
				&& $1/2$ & $0$ & $-2$ & $-2$ & $-4$ & $(M-2)/3$  \\
				&& $5/6$ & $0$ & $+2$ & $0$ & $+2$ & $(M+1)/3$  \\ \cline{2-8}
				& $\omega^2$ & $1/6$ & $-2$ & $-2$ & $0$ & $-4$ & $(M-2)/3$  \\
				&& $1/2$ & $-2$ & $+2$ & $+2$ & $+2$ & $(M+1)/3$  \\
				&& $5/6$ & $-2$ & $0$ & $-2$ & $-4$ & $(M-2)/3$  \\ \hline
				$6m+6$ & $1$ & $0$ & $+2$ & $+2$ & $+2$ & $+6$ & $M/3+1$  \\
				&& $1/3$ & $+2$ & $0$ & $-2$ & $0$ & $M/3$  \\
				&& $2/3$ & $+2$ & $-2$ & $0$ & $0$ & $M/3$  \\ \cline{2-8}
				& $\omega$ & $0$ & $0$ & $0$ & $0$ & $0$ & $M/3$  \\
				&& $1/3$ & $0$ & $-2$ & $+2$ & $0$ & $M/3$  \\
				&& $2/3$ & $0$ & $+2$ & $-2$ & $0$ & $M/3$  \\ \cline{2-8}
				& $\omega^2$ & $0$ & $-2$ & $-2$ & $-2$ & $-6$ & $M/3-1$  \\
				&& $1/3$ & $-2$ & $+2$ & $0$ & $0$ & $M/3$  \\
				&& $2/3$ & $-2$ & $0$ & $+2$ & $0$ & $M/3$  \\ \hline
			\end{tabular}
		}
	}
	\caption{The coefficients {$W_p$} of {the} delta functions in {Eqs.}\eqref{IndW31}-\eqref{IndW33} and {the} index {for the $T^2/{\mathbb{Z}}_3$ orbifold}.
		}
	\label{T5}
\end{table}

\subsection{$T^2/{\mathbb{Z}}_4$}
In the case of $N=4$, we have
\begin{align}
&\lim_{\rho \to \infty} {\rm{tr}}[\sigma_3 e^{{D}^2/\rho^2}]_{\eta} \notag \\
&=\frac{1}{4}\int_{T^2} d z d{\bar{z}}  \sum_{k=1}^{3} \sum_{m_k,n_k\in{\mathbb{Z}}} \eta^k(1-\omega^k)
e^{i\theta_{m_k,n_k}(z)}\delta^2(z-\omega^k z-m_k-n_k\tau).
\label{IndZ41}
\end{align}
The values of $(m_1,n_1)$, $(m_2,n_2)$ and $(m_3,n_3)$, which remain in the summations of Eq.\eqref{IndZ41} after the $z$-integration, are found to be
\begin{align}
(m_1,n_1)&=(0,0),(1,0), \\
(m_2,n_2)&=(0,0),(1,0),(0,1),(1,1), \\
(m_3,n_3)&=(0,0),(0,1). 
\end{align}
Then, it follows that Eq.\eqref{IndZ41} can be written into the form
\begin{align}
\lim_{\rho \to \infty} {\rm{tr}}[\sigma_3 e^{{D}^2/\rho^2}]_{\eta}
&=\frac{1}{2\times 4}\int_{T^2} dz d{\bar{z}} \sum_{p=1}^4 W_p \delta^2(z-{z_p}^f),
\label{Ind523}
\end{align}
where we have used the relations
\begin{align}
\delta^2(z-\omega z-m_1-n_1\tau)&=\frac{1}{2}\delta^2(z-(m_1-n_1)/{2}-(m_1+n_1)\tau/{2}), \\
\delta^2(z-\omega^2 z-m_2-n_2\tau)&=\frac{1}{4}\delta^2(z-m_2/2-n_2\tau/{2}), \\
\delta^2(z-\omega^3 z-m_3-n_3\tau)&=\frac{1}{2}\delta^2(z-(m_3+n_3)/{2}-(-m_3+n_3)\tau/{2}).
\end{align}
The $z_1^f$ and $z_2^f$ are the ${\mathbb{Z}}_4$ fixed points under the ${\mathbb{Z}}_4$ identification $z\sim \omega z$ $(\omega=e^{i2\pi/4}=i)$, i.e.\begin{align}
{\mathbb{Z}}_4\,\, {\rm{fixed \,\,point}}\, : \,\, z_1^f=0,\quad z_2^f=\frac{1}{2}(1+\tau).
\label{Z4fixed}
\end{align}
Since the ${\mathbb{Z}}_4$ group includes ${\mathbb{Z}}_2$ as its subgroup, there are additionally two {``${\mathbb{Z}}_2$ fixed points"}\footnote{{``${\mathbb{Z}}_2$ fixed points" are defined by
\begin{align}
z_I^f=\omega^2 z_I^f +m+n\tau \qquad {\rm{for}} \quad ^{\exists} m,n \,\in\,{\mathbb{Z}}.
\end{align}}} given by
\begin{align}
``{\mathbb{Z}}_2\,\, {\rm{fixed \,\,point}}"\, : \,\, z_3^f=\frac{1}{2},\quad z_4^f=\frac{\tau}{2}.
\label{Z2fixedsubgroup}
\end{align}
The coefficients $W_p\,\,(p=1,2,3,4)$ in Eq.\eqref{Ind523} are
\begin{align}
W_1&=\eta(1-\omega)e^{i\theta_{m_1,n_1}(z)}{\big{|}}^{(m_1,n_1)=(0,0)}_{z={z_1^f}} 
+\frac{1}{2}\eta^2(1-\omega^2)
e^{i\theta_{m_2,n_2}(z)}{\big{|}}^{(m_2,n_2)=(0,0)}_{z={z_1^f}} \notag \\
&\quad+\eta^3(1-\omega^3)
e^{i\theta_{m_3,n_3}(z)}{\big{|}}^{(m_3,n_3)=(0,0)}_{z={z_1^f}},
\label{IndW41}\\ 
W_2&=\eta(1-\omega)
e^{i\theta_{m_1,n_1}(z)}{\big{|}}^{(m_1,n_1)=(1,0)}_{z={z_2^f}}
+\frac{1}{2}\eta^2(1-\omega^2)
e^{i\theta_{m_2,n_2}(z)}{\big{|}}^{(m_2,n_2)=(1,1)}_{z={z_2^f}} \notag \\
&\quad+\eta^3(1-\omega^3)
e^{i\theta_{m_3,n_3}(z)}{\big{|}}^{(m_3,n_3)=(0,1)}_{z={z_2^f}}, \label{IndW42}\\
W_3&=\frac{1}{2}\eta^2(1-\omega^2)e^{i\theta_{m_2,n_2}(z)}{\big{|}}^{(m_2,n_2)=(1,0)}_{z={z_3^f}},\label{IndW43}
 \\
W_4&=\frac{1}{2}\eta^2(1-\omega^2)e^{i\theta_{m_2,n_2}(z)}{\big{|}}^{(m_2,n_2)=(0,1)}_{z={z_4^f}}.
\label{IndW44}
\end{align}
The explicit values of $W_p\,\,(p=1,2,3,4)$ are summarized in Table \ref{T6}.

\begin{table}[!ht]
	\centering
	{\tabcolsep = 3mm
		\renewcommand{\arraystretch}{1.2}
		\scalebox{0.80}{
			\begin{tabular}{ccc|c:c:c:c|c|c} \hline
			 flux& parity& twist & \multicolumn{4}{c|}{coefficients of the delta functions } &sum of coefficients & index\\
				$M$ & $\eta$ & $\alpha$ & $~~ W_1 ~~$ & $~~W_2 ~~$ & $~~W_3~~$ &$~~W_4~~$ & $\sum_{p=1}^4 W_p$ &
				$M/4+\sum_{p=1}^4 W_p/{8}$\\ \hline
				$4m+1$ & $1$ & $0$ & $+3$ & $+1$ &$+1$ & $+1$ & $+6$ & $(M+3)/4$ \\
				& & $1/2$ & $+3$ & $-3$ & $-1$ & $-1$ & $-2$  & $(M-1)/4$  \\ \cline{2-9}
				& $i$ & $0$ & $+1$ & $-1$ &$-1$ & $-1$ & $-2$ & $(M-1)/4$ \\
				& & $1/2$ & $+1$ & $+3$ & $+1$ & $+1$ & $+6$  & $(M+3)/4$  \\ \cline{2-9}
				 & $-1$ & $0$ & $-1$ & $-3$ &$+1$ & $+1$ & $-2$ & $(M-1)/4$ \\
				& & $1/2$ & $-1$ & $+1$ & $-1$ & $-1$ & $-2$  & $(M-1)/4$  \\ \cline{2-9}
				& $-i$ & $0$ & $-3$ & $+3$ &$-1$ & $-1$ & $-2$ & $(M-1)/4$ \\
				& & $1/2$ & $-3$ & $-1$ & $+1$ & $+1$ & $-2$  & $(M-1)/4$  \\ \hline	
				$4m+2$ & $1$ & $0$ & $+3$ & $-1$ &$+1$ & $+1$ & $+4$ & $(M+2)/4$ \\
				& & $1/2$ & $+3$ & $+3$ & $-1$ & $-1$ & $+4$  & $(M+2)/4$  \\ \cline{2-9}
				& $i$ & $0$ & $+1$ & $-3$ &$-1$ & $-1$ & $-4$ & $(M-2)/4$ \\
				& & $1/2$ & $+1$ & $+1$ & $+1$ & $+1$ & $+4$  & $(M+2)/4$  \\ \cline{2-9}
				 & $-1$ & $0$ & $-1$ & $+3$ &$+1$ & $+1$ & $+4$ & $(M+2)/4$ \\
				& & $1/2$ & $-1$ & $-1$ & $-1$ & $-1$ & $-4$  & $(M-2)/4$  \\ \cline{2-9}
				& $-i$ & $0$ & $-3$ & $+1$ &$-1$ & $-1$ & $-4$ & $(M-2)/4$ \\
				& & $1/2$ & $-3$ & $-3$ & $+1$ & $+1$ & $-4$  & $(M-2)/4$  \\ \hline	
				$4m+3$ & $1$ & $0$ & $+3$ & $-3$ &$+1$ & $+1$ & $+2$ & $(M+1)/4$ \\
				& & $1/2$ & $+3$ & $+1$ & $-1$ & $-1$ & $+2$  & $(M+1)/4$  \\ \cline{2-9}
				& $i$ & $0$ & $+1$ & $+3$ &$-1$ & $-1$ & $+2$ & $(M+1)/4$ \\
				& & $1/2$ & $+1$ & $-1$ & $+1$ & $+1$ & $+2$  & $(M+1)/4$  \\ \cline{2-9}
				 & $-1$ & $0$ & $-1$ & $+1$ &$+1$ & $+1$ & $+2$ & $(M+1)/4$ \\
				& & $1/2$ & $-1$ & $-3$ & $-1$ & $-1$ & $-6$  & $(M-3)/4$  \\ \cline{2-9}
				& $-i$ & $0$ & $-3$ & $-1$ &$-1$ & $-1$ & $-6$ & $(M-3)/4$ \\
				& & $1/2$ & $-3$ & $+3$ & $+1$ & $+1$ & $+2$  & $(M+1)/4$  \\ \hline	
				$4m+4$ & $1$ & $0$ & $+3$ & $+3$ &$+1$ & $+1$ & $+8$ & $M/4+1$ \\
				& & $1/2$ & $+3$ & $-1$ & $-1$ & $-1$ & $0$  & $M/4$  \\ \cline{2-9}
				& $i$ & $0$ & $+1$ & $+1$ &$-1$ & $-1$ & $0$ & $M/4$ \\
				& & $1/2$ & $+1$ & $-3$ & $+1$ & $+1$ & $0$  & $M/4$  \\ \cline{2-9}
				 & $-1$ & $0$ & $-1$ & $-1$ &$+1$ & $+1$ & $0$ & $M/4$ \\
				& & $1/2$ & $-1$ & $+3$ & $-1$ & $-1$ & $0$  & $M/4$  \\ \cline{2-9}
				& $-i$ & $0$ & $-3$ & $-3$ &$-1$ & $-1$ & $-8$ & $M/4-1$ \\
				& & $1/2$ & $-3$ & $+1$ & $+1$ & $+1$ & $0$  & $M/4$  \\ \hline							\end{tabular}
		}
	}
	\caption{The coefficients {$W_p$} of {the} delta functions in {Eqs.}\eqref{IndW41}-\eqref{IndW44} and {the} index {for the $T^2/{\mathbb{Z}}_4$ orbifold}.
			}
	\label{T6}
\end{table}

\subsection{$T^2/{\mathbb{Z}}_6$}
In the case of $N=6$, we obtain
\begin{align}
&\lim_{\rho \to \infty} {\rm{tr}}[\sigma_3 e^{{D}^2/\rho^2}]_{\eta} \notag \\
&=\frac{1}{6}\int_{T^2} d z d{\bar{z}}  \sum_{k=1}^{5}\sum_{m_k,n_k\in{\mathbb{Z}}}  \eta^k(1-\omega^k)
e^{i\theta_{m_k,n_k}(z)}\delta^2(z-\omega^k z-m_k-n_k\tau).
\label{IndZ61}
\end{align}
The values of $(m_1,n_1)$,$\cdots$, $(m_5,n_5)$, which remain in the summations of Eq.\eqref{IndZ61} after the $z$-integration, are given by
\begin{align}
(m_1,n_1)&=(0,0),\\
(m_2,n_2)&=(0,0),(1,0),(2,0), \\
(m_3,n_3)&=(0,0),(1,0),(0,1)(1,1),\\
 (m_4,n_4)&=(0,0),(0,1),(0,2), \\
 (m_5,n_5)&=(0,0).
\end{align}
Then, we can rewrite Eq.\eqref{IndZ61} into the form
\begin{align}
\lim_{\rho \to \infty} {\rm{tr}}[\sigma_3 e^{{D}^2/\rho^2}]_{\eta}
&=\frac{1}{2\times 6}\int_{T^2} dz d{\bar{z}} \sum_{p=1}^6 W_p \delta^2(z-{z_p^f}),
\label{IndZ6W}
\end{align}
where we have used the relations
\begin{align}
\delta^2(z-\omega z-m_1-n_1\tau)&=\delta^2(z+n_1-(m_1+n_1)\tau),\\
\delta^2(z-\omega^2 z-m_2-n_2\tau)&=\frac{1}{3}\delta^2(z-(m_2-n_2)/{3}-(m_2+2n_2)\tau/{3}), \\
\delta^2(z-\omega^3 z-m_3-n_3\tau)&=\frac{1}{4}\delta^2(z-m_3/{2}-n_3\tau/{2}),\\
\delta^2(z-\omega^4 z-m_4-n_4\tau)&=\frac{1}{3}\delta^2(z-(2m_4+n_4)/{3}-(-m_4+n_4)\tau/{3}), \\
\delta^2(z-\omega^5 z-m_5-n_5\tau)&=\delta^2(z-(m_5+n_5)+m_5\tau).
\end{align}
Note that there is one ${\mathbb{Z}}_6$ fixed point under the ${\mathbb{Z}}_6$ identification $z\sim \omega z$, i.e.
\begin{align}
{\mathbb{Z}}_6\,\, {\rm{fixed \,\,point}}\, : \,\, z_1^f=0.
\label{Z6fixed}
\end{align}
Since the ${\mathbb{Z}}_6$ group includes ${\mathbb{Z}}_3$ and ${\mathbb{Z}}_2$ as its subgroups, there are additionally two {``${\mathbb{Z}}_3$ fixed points"} and three {``${\mathbb{Z}}_2$ fixed points"}: 
\begin{align}
&``{\mathbb{Z}}_3\, {\rm{fixed\,\, point}}" :\quad
z_2^f=\frac{1}{3}(1+\tau),\quad z_3^f=\frac{2}{3}(1+\tau),\label{Z32fixedsubgroup1} \\
&``{\mathbb{Z}}_2\,{ \rm{fixed\,\, point} }":\quad
z_4^f=\frac{1}{2},\quad z_5^f=\frac{\tau}{2},\quad z_6^f=\frac{1}{2}(1+\tau).
\label{Z32fixedsubgroup}
\end{align}
The coefficients $W_p\,\,(p=1,2,\cdots,6)$ are given by
\begin{align}
W_1&=2\eta(1-\omega)
e^{i\theta_{m_1,n_1}(z)}{\big{|}}^{(m_1,n_1)=(0,0)}_{z={z_1^f}} 
+\frac{2}{3}\eta^2(1-\omega^2)
e^{i\theta_{m_2,n_2}(z)}{\big{|}}^{(m_2,n_2)=(0,0)}_{z={z_1^f}} 
\notag \\
&\quad+\frac{1}{2}\eta^3(1-\omega^3)
e^{i\theta_{m_3,n_3}(z)}{\big{|}}^{(m_3,n_3)=(0,0)}_{z={z_1^f}} 
+\frac{2}{3}\eta^4(1-\omega^4)
e^{i\theta_{m_4,n_4}(z)}{\big{|}}^{(m_4,n_4)=(0,0)}_{z={z_1^f}} 
\notag \\
&\quad
+2\eta^5(1-\omega^5)
e^{i\theta_{m_5,n_5}(z)}{\big{|}}^{(m_5,n_5)=(0,0)}_{z={z_1^f}} ,
\label{IndW61}\\ 
W_2&=\frac{2}{3}\eta^2(1-\omega^2)e^{i\theta_{m_2,n_2}(z)}{\big{|}}^{(m_2,n_2)=(1,0)}_{z={z_2^f}}
+\frac{2}{3}\eta^4(1-\omega^4)e^{i\theta_{m_4,n_4}(z)}{\big{|}}^{(m_4,n_4)=(0,1)}_{z={z_2^f}},  \label{IndW62}\\
W_3&=\frac{2}{3}\eta^2(1-\omega^2)e^{i\theta_{m_2,n_2}(z)}{\big{|}}^{(m_2,n_2)=(2,0)}_{z={z_3^f}}
+\frac{2}{3}\eta^4(1-\omega^4)e^{i\theta_{m_4,n_4}(z)}{\big{|}}^{(m_4,n_4)=(0,2)}_{z={z_3^f}}, \\
W_4&=\frac{1}{2}\eta^3(1-\omega^3)e^{i\theta_{m_3,n_3}(z)}{\big{|}}^{(m_3,n_3)=(1,0)}_{z={z_4^f}}, \label{IndW64}\\
W_5&=\frac{1}{2}\eta^3(1-\omega^3)e^{i\theta_{m_3,n_3}(z)}{\big{|}}^{(m_3,n_3)=(0,1)}_{z={z_5^f}}, \\
W_6&=\frac{1}{2}\eta^3(1-\omega^3)e^{i\theta_{m_3,n_3}(z)}{\big{|}}^{(m_3,n_3)=(1,1)}_{z={z_6^f}}.
\label{IndW66}
\end{align}
The explicit values of $W_p\,\,(p=1,2,\cdots,6)$ are summarized in Table \ref{T7}.

\begin{table}[!ht]
	\centering
	{\tabcolsep = 3mm
		\renewcommand{\arraystretch}{1.2}
		\scalebox{0.80}{
			\begin{tabular}{ccc|c:c:c:c:c:c|c|c} \hline
			 flux& parity& twist & \multicolumn{6}{c|}{coefficients of the delta functions } & sum of coefficients& index\\
				$M$ & $\eta$ & $\alpha$ & $W_1$ & $W_2$ & $W_3$ &$W_4$ &$W_5$ & $W_6$  & $\sum_{p=1}^6 W_p$ &$M/6+\sum_{p=1}^6 W_p/{12}$\\ \hline
				$6m+1$ & $1$ & $1/2$ & $+5$ & $-2$ &$-2$ & $-1$&$-1$ & $-1$ & $-2$ & $(M-1)/6$ \\
				 & $\omega$ & $1/2$ & $+3$ & $+2$ &$+2$ & $+1$&$+1$ & $+1$ & $+10$ & $(M+5)/6$ \\
				 & $\omega^2$ & $1/2$ & $+1$ & $0$ &$0$ & $-1$&$-1$ & $-1$ & $-2$ & $(M-1)/6$ \\
				 & $\omega^3$ & $1/2$ & $-1$ & $-2$ &$-2$ & $+1$&$+1$ & $+1$ & $-2$ & $(M-1)/6$ \\
				 & $\omega^4$ & $1/2$ & $-3$ & $+2$ &$+2$ & $-1$&$-1$ & $-1$ & $-2$ & $(M-1)/6$ \\
				 & $\omega^5$ & $1/2$ & $-5$ & $0$ &$0$ & $+1$&$+1$ & $+1$ & $-2$ & $(M-1)/6$ \\ \hline
				$6m+2$ & $1$ & $0$ & $+5$ & $0$ &$0$ & $+1$&$+1$ & $+1$ & $+8$ & $(M+4)/6$ \\
				 & $\omega$ & $0$ & $+3$ & $-2$ &$-2$ & $-1$&$-1$ & $-1$ & $-4$ & $(M-2)/6$ \\
				 & $\omega^2$ & $0$ & $+1$ & $+2$ &$+2$ & $+1$&$+1$ & $+1$ & $+8$ & $(M+4)/6$ \\
				 & $\omega^3$ & $0$ & $-1$ & $0$ &$0$ & $-1$&$-1$ & $-1$ & $-4$ & $(M-2)/6$ \\
				 & $\omega^4$ & $0$ & $-3$ & $-2$ &$-2$ & $+1$&$+1$ & $+1$ & $-4$ & $(M-2)/6$ \\
				 & $\omega^5$ & $0$ & $-5$ & $+2$ &$+2$ & $-1$&$-1$ & $-1$ & $-4$ & $(M-2)/6$ \\ \hline
				$6m+3$ & $1$ & $1/2$ & $+5$ & $+2$ &$+2$ & $-1$&$-1$ & $-1$ & $+6$ & $(M+3)/6$ \\
				 & $\omega$ & $1/2$ & $+3$ & $0$ &$0$ & $+1$&$+1$ & $+1$ & $+6$ & $(M+3)/6$ \\
				 & $\omega^2$ & $1/2$ & $+1$ & $-2$ &$-2$ & $-1$&$-1$ & $-1$ & $-6$ & $(M-3)/6$ \\
				 & $\omega^3$ & $1/2$ & $-1$ & $+2$ &$+2$ & $+1$&$+1$ & $+1$ & $+6$ & $(M+3)/6$ \\
				 & $\omega^4$ & $1/2$ & $-3$ & $0$ &$0$ & $-1$&$-1$ & $-1$ & $-6$ & $(M-3)/6$ \\
				 & $\omega^5$ & $1/2$ & $-5$ & $-2$ &$-2$ & $+1$&$+1$ & $+1$ & $-6$ & $(M-3)/6$ \\ \hline
				$6m+4$ & $1$ & $0$ & $+5$ & $-2$ &$-2$ & $+1$&$+1$ & $+1$ & $+4$ & $(M+2)/6$ \\
				 & $\omega$ & $0$ & $+3$ & $+2$ &$+2$ & $-1$&$-1$ & $-1$ & $+4$ & $(M+2)/6$ \\
				 & $\omega^2$ & $0$ & $+1$ & $0$ &$0$ & $+1$&$+1$ & $+1$ & $+4$ & $(M+2)/6$ \\
				 & $\omega^3$ & $0$ & $-1$ & $-2$ &$-2$ & $-1$&$-1$ & $-1$ & $-8$ & $(M-4)/6$ \\
				 & $\omega^4$ & $0$ & $-3$ & $+2$ &$+2$ & $+1$&$+1$ & $+1$ & $+4$ & $(M+2)/6$ \\
				 & $\omega^5$ & $0$ & $-5$ & $0$ &$0$ & $-1$&$-1$ & $-1$ & $-8$ & $(M-4)/6$ \\ \hline
				$6m+5$ & $1$ & $1/2$ & $+5$ & $0$ &$0$ & $-1$&$-1$ & $-1$ & $+2$ & $(M+1)/6$ \\
				 & $\omega$ & $1/2$ & $+3$ & $-2$ &$-2$ & $+1$&$+1$ & $+1$ & $+2$ & $(M+1)/6$ \\
				 & $\omega^2$ & $1/2$ & $+1$ & $+2$ &$+2$ & $-1$&$-1$ & $-1$ & $+2$ & $(M+1)/6$ \\
				 & $\omega^3$ & $1/2$ & $-1$ & $0$ &$0$ & $+1$&$+1$ & $+1$ & $+2$ & $(M+1)/6$ \\
				 & $\omega^4$ & $1/2$ & $-3$ & $-2$ &$-2$ & $-1$&$-1$ & $-1$ & $-10$ & $(M-5)/6$ \\
				 & $\omega^5$ & $1/2$ & $-5$ & $+2$ &$+2$ & $+1$&$+1$ & $+1$ & $+2$ & $(M+1)/6$ \\ \hline
				$6m+6$ & $1$ & $0$ & $+5$ & $+2$ &$+2$ & $+1$&$+1$ & $+1$ & $+12$ & $M/6+1$ \\
				 & $\omega$ & $0$ & $+3$ & $0$ &$0$ & $-1$&$-1$ & $-1$ & $0$ & $M/6$ \\
				 & $\omega^2$ & $0$ & $+1$ & $-2$ &$-2$ & $+1$&$+1$ & $+1$ & $0$ & $M/6$ \\
				 & $\omega^3$ & $0$ & $-1$ & $+2$ &$+2$ & $-1$&$-1$ & $-1$ & $0$ & $M/6$ \\
				 & $\omega^4$ & $0$ & $-3$ & $0$ &$0$ & $+1$&$+1$ & $+1$ & $0$ & $M/6$ \\
				 & $\omega^5$ & $0$ & $-5$ & $-2$ &$-2$ & $-1$&$-1$ & $-1$ & $-12$ & $M/6-1$ \\ \hline
				\end{tabular}
		}
	}
	\caption{The coefficients {$W_p$} of {the} delta functions in {Eqs.}\eqref{IndW61}-\eqref{IndW66} and {the} index {for the $T^2/{\mathbb{Z}}_6$ orbifold} .
		}
	\label{T7}
\end{table}

\section{Winding numbers at fixed points on  $T^2/{\mathbb{Z}}_N$ orbifolds}
In this section, we introduce winding numbers at the fixed points on the $T^2/{\mathbb{Z}}_N$ $(N=2,3,4,6)$ orbifolds, and clarify the relation between the winding numbers and the coefficients $W_p$ in front of the delta functions in Eqs.\eqref{IndWp}, \eqref{Ind512}, \eqref{Ind523} and \eqref{IndZ6W}.

Let us define the winding number for the ${\mathbb{Z}}_N$ eigenstate $\psi_{T^2/{\mathbb{Z}}_N+,n}^{(j+\alpha_1,\alpha_2)}(z)_{\eta}$ at a fixed point $z_p^f$ as
\begin{align}
\chi_p(\eta,(\alpha_1,\alpha_2),M)\equiv \frac{1}{2\pi i}\oint_{C_p} d {\bm{l}} \cdot  \nabla \log (\psi_{T^2/{\mathbb{Z}}_N+,n}^{(j+\alpha_1,\alpha_2)}(z)_{\eta}),
\end{align}
where $C_p$ denotes a sufficiently small circle encircled anticlockwise around the fixed point $z=z_p^f$. The line integral along the contour $C_p$ gives a winding number, i.e. how many times $\psi_{T^2/{\mathbb{Z}}_N+,n}^{(j+\alpha_1,\alpha_2)}(z)_{\eta}$ wraps around the origin. It should be noted that the winding numbers at fixed points are determined only by the  ${\mathbb{Z}}_N$ transformation \eqref{ZNes1} and the boundary conditions \eqref{BCs}.

We define the winding numbers $\chi_{\pm p}$ around the ${\mathbb{Z}}_N$ fixed points $z_p^f$ as\footnote{We may define $\chi_{-p}$ as~\cite{Sakamoto:2020vdy}
\begin{align}
\chi_{-p}\equiv \chi_p({\bar{\omega}}{\bar{\eta}},(-\alpha_1,-\alpha_2),-M).
\end{align}
It is interesting to note that Eqs.\eqref{WN1} and \eqref{WN2} turn out to lead to the same results, as verified by explicit computations.}

\begin{align}
&\chi_{+p}\equiv \chi_p(\eta,(\alpha_1,\alpha_2),M), \label{WN1}\\
&\chi_{-p}\equiv -\chi_p(\omega \eta,(\alpha_1,\alpha_2),M)+N.
\label{WN2}
\end{align}

Let us examine the winding numbers $\chi_{\pm p}$ around the fixed point $z_p^f$. From Eqs.\eqref{BCs} and \eqref{ZNes1}, we can show that $\psi_{T^2/{\mathbb{Z}}_N+,n}^{(j+\alpha_1,\alpha_2)}(z)_{\eta}$ satisfies the relation
\begin{align}
\psi_{T^2/{\mathbb{Z}}_N+,n}^{(j+\alpha_1,\alpha_2)}(z_p^f+\omega Z)_{\eta}&=e^{i(mq\Lambda_1(\omega Z)+nq\Lambda_2(\omega Z))}\eta e^{im(q\Lambda_1(z_p^f)+2\pi\alpha_1)+in(q\Lambda_2(z_p^f)+2\pi \alpha_2)+i\pi M mn} \notag \\
&\qquad \times \psi_{T^2/{\mathbb{Z}}_N+,n}^{(j+\alpha_1,\alpha_2)}(z_p^f+ Z)_{\eta}.
\label{WDdefine}
\end{align}
It follows that by taking the limit of $Z\to 0$, the winding numbers $\chi_{\pm p}$ around the ${\mathbb{Z}}_{N}$ fixed point $z_p^f$ are found to be
\begin{align}
\chi_{+p}&=N\left(
\frac{mq}{2\pi}\Lambda_1(z_p^f)+m\alpha_1+\frac{nq}{2\pi}\Lambda_2(z_p^f)+n \alpha_2+\frac{1}{2}Mmn\right)+l  \qquad {\rm{mod}} \,\,N,\label{WDN1}\\
\chi_{-p}&=-N \left(
\frac{mq}{2\pi}\Lambda_1(z_p^f)+m\alpha_1+\frac{nq}{2\pi}\Lambda_2(z_p^f)+n \alpha_2+\frac{1}{2}Mmn\right)-l-1+N  \qquad {\rm{mod}}\,\, N,
\label{WDN}
\end{align}
where $\eta=\omega^l\,(l=0,1,\cdots, N-1)$. Since the winding numbers $\chi_{\pm p}$ can be obtained from Eq.\eqref{WDdefine} up to mod $N$ around the ${\mathbb{Z}}_{N}$ fixed point $z_p^f$, we restrict the values of $\chi_{\pm p}$ to $0,1,\cdots,N-1$.

A crucial observation is that the coefficient $W_p$ can be expressed, in terms of the winding numbers $\chi_{\pm p}$, as\footnote{Note that in Eqs.\eqref{WN1}$-$\eqref{WDderive}, $N$ should take the value 2 for the $\mathbb{Z}_2$ fixed points on the $T^2/\mathbb{Z}_4$ and $T^2/\mathbb{Z}_6$ orbifolds, and 3 for the $\mathbb{Z}_3$ fixed points on the $T^2/\mathbb{Z}_6$ orbifold.}
\begin{align}
W_p=-2\chi_{+p}+N-1=-\chi_{+p}+\chi_{-p}.
\label{WDderive}
\end{align}
The proof of Eq.\eqref{WDderive} is given in Appendix C.

By use of the relation \eqref{WDderive}, we can express $\lim_{\rho \to \infty} {\rm{tr}}[\sigma_3 e^{{D}^2/\rho^2}]_{\eta}$ into the form
\begin{align}
\lim_{\rho \to \infty} {\rm{tr}}[\sigma_3 e^{{D}^2/\rho^2}]_{\eta}
=\frac{1}{2N}\int_{T^2}dz d\bar{z} \sum_{p}W_p\delta^2(z-z_p^f)
=\frac{1}{2N}(-V_+ +V_-),
\label{Indc}
\end{align}
where $V_{\pm}$ are the sums of the winding numbers $\chi_{\pm p}$ around the fixed points $z_p^f$, i.e.
\begin{align}
V_{\pm}=\sum_{p} \chi_{\pm p}.
\end{align}
Therefore, we conclude that the index formula on the $T^2/{\mathbb{Z}}_N$ $(N=2,3,4,6)$ orbifold with magnetic flux is given by
\begin{align}
n_+-n_-=\frac{M}{N}-\frac{V_+ -V_-}{2N}.
\label{IndexZN}
\end{align}
If we use the relation $\chi_{-p}=-\chi_{+p}+N-1$ for the ${\mathbb{Z}}_N$ fixed point $z_p^f$, the formula \eqref{IndexZN} can be rewritten as
\begin{align}
n_+-n_-={\frac{M-V_{+}}{N}+1},
\label{IndexZN1}
\end{align}
which is the relation found in {Ref.}~\cite{Sakamoto:2020pev}.

\section{Conclusion and discussion}
In this paper, we have derived the index 
\begin{align}
n_+-n_-=\frac{M}{N}+\frac{1}{2N}(-V_+ +V_-)=\frac{M-V_+}{N}+1
\label{IndexZNend}
\end{align}
on the $T^2/{\mathbb{Z}}_N$ orbifolds with magnetic flux by using the trace formula. 
The first term {$M/N$} of Eq.\eqref{IndexZNend} can be evaluated by the Fujikawa method, where the factor $1/N$ comes from the fact {that} the area of the $T^2/{\mathbb{Z}}_N$ orbifold is $1/N$ of that of the torus $T^2$. {Our main subject of this paper is to derive the nontrivial terms of $(-V_{+}+V_{-})/2N$ (or $-V_{+}/N+1$) in Eq.\eqref{IndexZNend} from the trace formula.}

{To derive the index $n_{+}-n_{-}$, we have used the trace formula \eqref{Ind} in this paper. By constructing the complete set of the mode functions on $T^2/{\mathbb{Z}}_2$ in terms of those on $T^2$, we have succeeded in evaluating the first term on the right-hand side of Eq.\eqref{Ind2} for the $T^2/{\mathbb{Z}}_2$ orbifold. We have then found that the first term of Eq.\eqref{Ind2} is determined only by the information at the fixed points on $T^2/{\mathbb{Z}}_2$, as shown in Eq.\eqref{IndWp}, and further that the coefficients $W_p$ of the delta functions $\delta^2(z-z_p^f)$ are related to the winding numbers $-\chi_{+p}+\chi_{-p}$ (or $-2\chi_{+p}+1$) at the fixed points $z_p^f$ (see Eq.\eqref{WDderive}). We finally arrived at the index formula given by Eq.\eqref{IndexZN} (or Eq.\eqref{IndexZN1}).}

{Although it was not possible to construct complete sets of mode functions on the $T^2/{\mathbb{Z}}_N$ $(N=3,4,6)$ orbifolds for arbitrary $M$, we were able to get the desired result \eqref{IndexZN} starting from the expression \eqref{Ind57}. Thus, we confirmed the zero-mode counting formula \eqref{Zeromodecountingformula} given in Ref.~\cite{Sakamoto:2020pev} as an {expression of the Atiyah-Singer index theorem}. It is interesting to note that $M/N$ and $-V_{+}/N+1~~$  (or  $(-V_{+}+V_{-})/(2N)$) are not always integer-valued, but the sum of them $(M-V_{+})/N+1~$ (or $(M/N+(-V_{+}+V_{-})/(2N)$) becomes an integer in any case. Therefore, the combination is nontrivial in the index theorem point of view.} 

{Some works remain to be done. The first one is to verify the relation \eqref{Ind57} for arbitrary $M$. To this end, one might try to construct complete sets of mode functions on the $T^2/{\mathbb{Z}}_N$ $(N=3,4,6)$ orbifolds, but it seems to be hard since the ${\mathbb{Z}}_N$ transformation of the mode functions on $T^2$ is quite complicated. Some better ideas will be needed.}

{The second one is to clarify physical or geometrical roles of the terms $-V_{+}/N+1$ (or $(-V_{+}+V_{-})/(2N)$)  from the viewpoint of the Atiyah-Singer index theorem. Since the $T^2/{\mathbb{Z}}_N$ orbifolds have singularities at the fixed points, we cannot apply the Atiyah-Singer index theorem directly to the orbifolds. If one can replace the singularities on the orbifolds with some parts of smooth manifolds, we could obtain smooth blow-up manifolds instead of the orbifolds. Then, we can apply the Atiyah-Singer index theorem to the manifolds without singularities and are expected to reveal physical or geometrical meanings of the terms $\,-V_{+}/N+1~$ (or $(-V_{+}+V_{-})/(2N)$). The work along this line {is} found in Ref.~\cite{Kobayashi:2022tti,Kobayashi:2022xsk}.}

An interesting extension of this work is to higher dimensions. In the two-dimensional case, we have found that the winding number appears as a topological invariant in the index theorem. In higher dimensions, we anticipate that other topological objects besides the winding number contribute to it. We will research the extension to higher dimensions elsewhere.

\section*{Acknowledgment}
M.T. was supported by Grant-in-Aid for Japan Society for the Promotion of Science (JSPS) Research Fellow and JSPS KAKENHI Grant Number JP 21J20739. M.S. was supported by JSPS KAKENHI Grant Number JP 18K03649. Y.T. was supported in part by Scuola Normale, by INFN (IS GSS-Pi) and by MIUR-PRIN contract 2017CC72MK\_003.
\appendix
\renewcommand{\thesection}{\Alph{section}}

{\section{Gamma matrices}
The 6d Gamma matrices are taken as
 \begin{gather}
	\{ \Gamma^K, \Gamma^L \} = -2 \eta^{KL} \qquad (K,L = 0, 1, 2, 3, 5, 6), \\
	\eta^{KL} = {\rm diag} \, (-1, 1, 1, 1, 1, 1), \\
	\Gamma^\mu = 
	\begin{pmatrix}
		\gamma^\mu & 0 \\
		0 & \gamma^\mu
	\end{pmatrix} \qquad (\mu = 0, 1, 2, 3), \\
	\Gamma^5 = 
	\begin{pmatrix}
		0 & i\gamma_5 \\
		i\gamma_5 & 0
	\end{pmatrix}, \qquad 
	\Gamma^6 = 
	\begin{pmatrix}
		0 & \gamma_5 \\
		-\gamma_5 & 0
	\end{pmatrix}, \qquad 
	\Gamma_7 = 
	\begin{pmatrix}
		\gamma_5 & 0 \\
		0 & - \gamma_5
	\end{pmatrix}.
\end{gather}}

\section{The proof of Eq.\eqref{Ind3}}
{In this Appendix, we show Eq.\eqref{Ind3} only for the case of $(\alpha_1,\alpha_2)=(0,0)$ with $M=\rm{even/odd}$  and $\eta=\pm 1$.}
 
\subsubsection*{\underline{$(\alpha_1,\alpha_2)=(0,0),\,M=2m+1,\,\eta=+1$}}
Let us first consider the case of $(\alpha_1,\alpha_2)=(0,0),\,M=2m+1$ and $\eta=+1$. We start with
\begin{align}
I_{\eta=+1}{\equiv}
\sum_{n=0}^{\infty} \sum_{j}
\psi_{T^2/{\mathbb{Z}}_2+,n}^{(j,0)}(z)_{+1}[\psi_{T^2/{\mathbb{Z}}_2+,n}^{(j,0)}(z)_{+1}]^{\ast}-\sum_{n=1}^{\infty} \sum_{j}
\psi_{T^2/{\mathbb{Z}}_2-,n}^{(j,0)}(z)_{+1}[\psi_{T^2/{\mathbb{Z}}_2-,n}^{(j,0)}(z)_{+1}]^{\ast}.
\label{Index1}
\end{align}
From Table \ref{T1} and Eqs.\eqref{Z2KKmode1} and \eqref{Z2KKmode}, the label $j$ for $\psi_{T^2/{\mathbb{Z}}_2\pm,n}^{(j,0)}(z)_{+1}$ runs from 0 (1) to $\tfrac{M-1}{2}$ ($\tfrac{M-1}{2}$) for $n={\rm{even}}$ (odd). Thus, $I_{\eta=+1}$ can be represented as
\begin{align}
I_{\eta=+1}=
&\sum_{k=0}^{\infty} \sum_{j=0}^{\frac{M-1}{2}}
\psi_{T^2/{\mathbb{Z}}_2+,2k}^{(j,0)}(z)_{+1}[\psi_{T^2/{\mathbb{Z}}_2+,2k}^{(j,0)}(z)_{+1}]^{\ast} \notag \\
+&\sum_{k=0}^{\infty} \sum_{j=1}^{\frac{M-1}{2}}
\psi_{T^2/{\mathbb{Z}}_2+,2k+1}^{(j,0)}(z)_{+1}[\psi_{T^2/{\mathbb{Z}}_2+,2k+1}^{(j,0)}(z)_{+1}]^{\ast} \notag \\
-&\sum_{k=0}^{\infty}\sum_{j=0}^{\frac{M-1}{2}}
\psi_{T^2/{\mathbb{Z}}_2-,2k+2}^{(j,0)}(z)_{+1}[\psi_{T^2/{\mathbb{Z}}_2-,2k+2}^{(j,0)}(z)_{+1}]^{\ast} \notag \\
-&\sum_{k=0}^{\infty} \sum_{j=1}^{\frac{M-1}{2}}
\psi_{T^2/{\mathbb{Z}}_2-,2k+1}^{(j,0)}(z)_{+1}[\psi_{T^2/{\mathbb{Z}}_2-,2k+1}^{(j,0)}(z)_{+1}]^{\ast}.
\label{Index2}
\end{align}

An important observation is that we can add $j=0$ terms in the second and the forth terms of Eq.\eqref{Index2} because
\begin{align}
\psi_{T^2/{\mathbb{Z}}_2+,0}^{(j,0)}(z)_{-1}=
{\cal{N}}_{+,-1}^{(j,0)}\Big[\psi_{T^2+,0}^{(j,0)}(z)-\psi_{T^2+,0}^{(j,0)}(-z)\Big]
\end{align}
identically vanishes for $j=0$. Then, taking ${\cal{N}}_{+,-1}^{(0,0)}=1/\sqrt{2}$, we can rewrite Eq.\eqref{Index2} as follows:
\begin{align}
I_{\eta=+1}&=
\sum_{n=0}^{\infty} \sum_{j=0}^{\frac{M-1}{2}}
\psi_{T^2/{\mathbb{Z}}_2+,n}^{(j,0)}(z)_{+1}[\psi_{T^2/{\mathbb{Z}}_2+,n}^{(j,0)}(z)_{+1}]^{\ast} 
 -\sum_{n=1}^{\infty} \sum_{j=0}^{\frac{M-1}{2}}
\psi_{T^2/{\mathbb{Z}}_2-,n}^{(j,0)}(z)_{+1}[\psi_{T^2/{\mathbb{Z}}_2-,n}^{(j,0)}(z)_{+1}]^{\ast}
\notag \\
&=
\sum_{n=0}^{\infty} \sum_{j=0}^{\frac{M-1}{2}} \sum_{l=0}^{1} \sum_{l^{\prime}=0}^{1}
(1-\omega^{-l+l^{\prime}})|{\cal{N}}_{+,+1}^{(j,0)}|^2
\psi_{T^2+,n}^{(j,0)}(\omega^l z)[\psi_{T^2+,n}^{(j,0)}(\omega^{l^{\prime}}z)]^{\ast} \notag \\
&=\sum_{n=0}^{\infty} \sum_{j=0}^{\frac{M-1}{2}} 
(1-\omega)|{\cal{N}}_{+,+1}^{(j,0)}|^2
\Big{\{}\psi_{T^2+,n}^{(j,0)}(z)[\psi_{T^2+,n}^{(j,0)}(\omega z)]^{\ast}+
\psi_{T^2+,n}^{(j,0)}(\omega z)[\psi_{T^2+,n}^{(j,0)}( z)]^{\ast}\Big{\}} ,
\label{Index3} 
\end{align}
where we have used the relations \eqref{Z2KKmode1}, \eqref{Z2KKmode} and ${\cal{N}}_{+,+1}^{(j,0)}={\cal{N}}_{+,-1}^{(j,0)}$ for $j=0,1,\cdots,\tfrac{M-1}{2}$ in the second equality, and $\omega=-1$ in the last equality.

Using the relation \eqref{Torusz2} and $\psi_{T^2+,n}^{(M,0)}(z)=\psi_{T^2+,n}^{(0,0)}(z)$, we can further rewrite Eq.\eqref{Index3} as
\begin{align}
I_{\eta=+1}
&=\sum_{n=0}^{\infty}(1-\omega){\Big\{}2 |{\cal{N}}_{+,+1}^{(0,0)}|^2
\psi_{T^2+,n}^{(0,0)}(z)[\psi_{T^2+,n}^{(0,0)}(\omega z)]^{\ast} \notag \\
&\quad+
 \sum_{j=1}^{\frac{M-1}{2}} |{\cal{N}}_{+,+1}^{(j,0)}|^2
(\psi_{T^2+,n}^{(j,0)}(z)[\psi_{T^2+,n}^{(j,0)}(\omega z)]^{\ast}+
\psi_{T^2+,n}^{(M-j,0)}( z)[\psi_{T^2+,n}^{(M-j,0)}( \omega z)]^{\ast}) {\Big\}}
\notag \\
&=\sum_{n=0}^{\infty} \sum_{j=0}^{M-1}(1-\omega)
\psi_{T^2+,n}^{(j,0)}(z)[\psi_{T^2+,n}^{(j,0)}(\omega z)]^{\ast},
\label{Index4} 
\end{align}
where we have used ${\cal{N}}_{+,+1}^{(0,0)}=1/\sqrt{2}$ and ${\cal{N}}_{+,+1}^{(j,0)}=1$ $(j=1,2,\cdots,\tfrac{M-1}{2})$ for $M={\rm{odd}}$ from Table \ref{T2}.

Taking the limit of $z^{\prime} \to z$ and $\rho \to \infty$ in Eq.\eqref{Indz2} and then substituting Eq.\eqref{Index4} into Eq.\eqref{Indz2}, we finally arrive at
\begin{align}
\lim_{\rho \to \infty} {\rm{tr}}[\sigma_3 e^{{D}^2/\rho^2}]_{\eta=+1} 
=\frac{1}{2}\int_{T^2} dz d{\bar{z}} \sum_{n=0}^{\infty} \sum_{j=0}^{M-1}  (1-\omega)
 \psi^{(j,0)}_{T^2+,n}(z)[ \psi^{(j,0)}_{T^2+,n}(\omega z)]^{\ast},
\label{Index5}
\end{align}
where we have used the fact that $\int_{T^2/{\mathbb{Z}}_2} dz d{\bar{z}}$ can be replaced by $\tfrac{1}{2}\int_{T^2} dz d{\bar{z}}$ in the final expression. Thus, we have succeeded in verifying Eq.\eqref{Ind3} for $(\alpha_1,\alpha_2)=(0,0)$, $M={\rm{odd}}$ and $\eta=+1$.

\subsubsection*{\underline{$(\alpha_1,\alpha_2)=(0,0),\,M=2m+1,\,\eta=-1$}}
We next discuss the case of $(\alpha_1,\alpha_2)=(0,0),\,M=2m+1$ and $\eta=-1$. Let us consider
\begin{align}
I_{\eta=-1}\equiv
\sum_{n=0}^{\infty} \sum_{j}
\psi_{T^2/{\mathbb{Z}}_2+,n}^{(j,0)}(z)_{-1}[\psi_{T^2/{\mathbb{Z}}_2+,n}^{(j,0)}(z)_{-1}]^{\ast}-\sum_{n=1}^{\infty} \sum_{j}
\psi_{T^2/{\mathbb{Z}}_2-,n}^{(j,0)}(z)_{-1}[\psi_{T^2/{\mathbb{Z}}_2-,n}^{(j,0)}(z)_{-1}]^{\ast}.
\label{Index2_1}
\end{align}
From Table \ref{T1} and Eqs.\eqref{Z2KKmode1} and \eqref{Z2KKmode}, the label $j$ for $\psi_{T^2/{\mathbb{Z}}_2\pm,n}^{(j,0)}(z)_{-1}$ runs from 1 (0) to $\tfrac{M-1}{2}$ ($\tfrac{M-1}{2}$) for $n={\rm{even}}$ (odd). Thus, $I_{\eta=-1}$ can be expressed as
\begin{align}
I_{\eta=-1}=
&\sum_{k=0}^{\infty} \sum_{j=1}^{\frac{M-1}{2}}
\psi_{T^2/{\mathbb{Z}}_2+,2k}^{(j,0)}(z)_{-1}[\psi_{T^2/{\mathbb{Z}}_2+,2k}^{(j,0)}(z)_{-1}]^{\ast} \notag \\
+&\sum_{k=0}^{\infty} \sum_{j=0}^{\frac{M-1}{2}}
\psi_{T^2/{\mathbb{Z}}_2+,2k+1}^{(j,0)}(z)_{-1}[\psi_{T^2/{\mathbb{Z}}_2+,2k+1}^{(j,0)}(z)_{-1}]^{\ast} \notag \\
-&\sum_{k=0}^{\infty}\sum_{j=1}^{\frac{M-1}{2}}
\psi_{T^2/{\mathbb{Z}}_2-,2k+2}^{(j,0)}(z)_{-1}[\psi_{T^2/{\mathbb{Z}}_2-,2k+2}^{(j,0)}(z)_{-1}]^{\ast} \notag \\
-&\sum_{k=0}^{\infty} \sum_{j=0}^{\frac{M-1}{2}}
\psi_{T^2/{\mathbb{Z}}_2-,2k+1}^{(j,0)}(z)_{-1}[\psi_{T^2/{\mathbb{Z}}_2-,2k+1}^{(j,0)}(z)_{-1}]^{\ast}.
\label{Index2_2}
\end{align}

Since $\psi_{T^2/{\mathbb{Z}}_2\pm,2k}^{(j,0)}(z)_{-1}=0$ for $j=0$, we can add $j=0$ terms in the first and the third terms of Eq.\eqref{Index2_2}. Then, we find
\begin{align}
I_{\eta=-1}&=
\sum_{n=0}^{\infty} \sum_{j=0}^{\frac{M-1}{2}}
\psi_{T^2/{\mathbb{Z}}_2+,n}^{(j,0)}(z)_{-1}[\psi_{T^2/{\mathbb{Z}}_2+,n}^{(j,0)}(z)_{-1}]^{\ast} 
 -\sum_{n=1}^{\infty} \sum_{j=0}^{\frac{M-1}{2}}
\psi_{T^2/{\mathbb{Z}}_2-,n}^{(j,0)}(z)_{-1}[\psi_{T^2/{\mathbb{Z}}_2-,n}^{(j,0)}(z)_{-1}]^{\ast}
\notag \\
&=
\sum_{n=0}^{\infty} \sum_{j=0}^{\frac{M-1}{2}} \sum_{l=0}^{1} \sum_{l^{\prime}=0}^{1}\eta^{-l+l^{\prime}}(1{-}\omega^{-l+l^{\prime}})
|{\cal{N}}_{+,+1}^{(j,0)}|^2
\psi_{T^2+,n}^{(j,0)}(\omega^l z)[\psi_{T^2+,n}^{(j,0)}(\omega^{l^{\prime}}z)]^{\ast} \notag \\
&=\sum_{n=0}^{\infty} \sum_{j=0}^{\frac{M-1}{2}}
\eta(1-\omega) |{\cal{N}}_{+,+1}^{(j,0)}|^2
{\Big\{}\psi_{T^2+,n}^{(j,0)}(z)[\psi_{T^2+,n}^{(j,0)}(\omega z)]^{\ast}+
\psi_{T^2+,n}^{(j,0)}(\omega z)[\psi_{T^2+,n}^{(j,0)}( z)]^{\ast}{\Big\}} ,
\label{Index2_3} 
\end{align}
where we have used the relations \eqref{Z2KKmode1}, \eqref{Z2KKmode} and ${\cal{N}}_{+,+1}^{(j,0)}={\cal{N}}_{+,-1}^{(j,0)}$ for $j=0,1,\cdots,\tfrac{M-1}{2}$ in the second equality, and $\eta=\omega=-1$ in the third equality.

Using the relation \eqref{Torusz2} and $\psi_{T^2+,n}^{(M,0)}(z)=\psi_{T^2+,n}^{(0,0)}(z)$, we can further rewrite Eq.\eqref{Index2_3} as
\begin{align}
I_{\eta=-1}
&=\sum_{n=0}^{\infty}\eta(1-\omega){\Bigl\{}2 |{\cal{N}}_{+,+1}^{(0,0)}|^2
\psi_{T^2+,n}^{(0,0)}(z)[\psi_{T^2+,n}^{(0,0)}(\omega z)]^{\ast} \notag \\
&\quad+
 \sum_{j=1}^{\frac{M-1}{2}} |{\cal{N}}_{+,+1}^{(j,0)}|^2
(\psi_{T^2+,n}^{(j,0)}(z)[\psi_{T^2+,n}^{(j,0)}(\omega z)]^{\ast}+
\psi_{T^2+,n}^{(M-j,0)}( z)[\psi_{T^2+,n}^{(M-j,0)}( \omega z)]^{\ast}) {\Bigr\}}
\notag \\
&=\sum_{n=0}^{\infty} \sum_{j=0}^{M-1}\eta(1-\omega)
\psi_{T^2+,n}^{(j,0)}(z)[\psi_{T^2+,n}^{(j,0)}(\omega z)]^{\ast},
\label{Index2_4} 
\end{align}
where we have used ${\cal{N}}_{+,+1}^{(0,0)}=1/\sqrt{2}$ and ${\cal{N}}_{+,+1}^{(j,0)}=1$ $(j=1,2,\cdots,\tfrac{M-1}{2})$ for $M={\rm{odd}}$ from Table \ref{T2}.

Taking the limit of $z^{\prime} \to z$ and $\rho \to \infty$ in Eq.\eqref{Indz2} and then substituting Eq.\eqref{Index2_4} into Eq.\eqref{Indz2}, we find that Eq.\eqref{Ind3} holds for $(\alpha_1,\alpha_2)=(0,0),\,M={\rm{odd}}$ and $\eta=-1$.
We have succeeded in verifying Eq.\eqref{Ind3} for the case of $(\alpha_1,\alpha_2)=(0,0),\,M={\rm{odd}}$ and $\eta=\pm 1$.

\subsubsection*{\underline{$(\alpha_1,\alpha_2)=(0,0),\,M=2m+2,\,\eta=+1$}}
Let us consider the case of $(\alpha_1,\alpha_2)=(0,0),\,M=2m+2$ and $\eta=+1$. 
\begin{align}
I_{\eta=+1}{\equiv}
\sum_{n=0}^{\infty} \sum_{j}
\psi_{T^2/{\mathbb{Z}}_2+,n}^{(j,0)}(z)_{+1}[\psi_{T^2/{\mathbb{Z}}_2+,n}^{(j,0)}(z)_{+1}]^{\ast}-\sum_{n=1}^{\infty} \sum_{j}
\psi_{T^2/{\mathbb{Z}}_2-,n}^{(j,0)}(z)_{+1}[\psi_{T^2/{\mathbb{Z}}_2-,n}^{(j,0)}(z)_{+1}]^{\ast}.
\label{Indexeven1}
\end{align}
From Table \ref{T1} and Eqs.\eqref{Z2KKmode1} and \eqref{Z2KKmode}, the label $j$ for $\psi_{T^2/{\mathbb{Z}}_2\pm,n}^{(j,0)}(z)_{+1}$ runs from 0 (1) to $\tfrac{M}{2}$ ($\tfrac{M}{2}-1$) for $n={\rm{even}}$ (odd). Thus, $I_{\eta=+1}$ can be represented as
\begin{align}
I_{\eta=+1}=
&\sum_{k=0}^{\infty} \sum_{j=0}^{\frac{M}{2}}
\psi_{T^2/{\mathbb{Z}}_2+,2k}^{(j,0)}(z)_{+1}[\psi_{T^2/{\mathbb{Z}}_2+,2k}^{(j,0)}(z)_{+1}]^{\ast} \notag \\
+&\sum_{k=0}^{\infty} \sum_{j=1}^{\frac{M}{2}-1}
\psi_{T^2/{\mathbb{Z}}_2+,2k+1}^{(j,0)}(z)_{+1}[\psi_{T^2/{\mathbb{Z}}_2+,2k+1}^{(j,0)}(z)_{+1}]^{\ast} \notag \\
-&\sum_{k=0}^{\infty}\sum_{j=0}^{\frac{M}{2}}
\psi_{T^2/{\mathbb{Z}}_2-,2k+2}^{(j,0)}(z)_{+1}[\psi_{T^2/{\mathbb{Z}}_2-,2k+2}^{(j,0)}(z)_{+1}]^{\ast} \notag \\
-&\sum_{k=0}^{\infty} \sum_{j=1}^{\frac{M}{2}-1}
\psi_{T^2/{\mathbb{Z}}_2-,2k+1}^{(j,0)}(z)_{+1}[\psi_{T^2/{\mathbb{Z}}_2-,2k+1}^{(j,0)}(z)_{+1}]^{\ast}.
\label{Indexeven2}
\end{align}

Since $\psi_{T^2/{\mathbb{Z}}_2\pm,2k+1}^{(j,0)}(z)_{-1}=0$ for $j=0,\tfrac{M}{2}$, we can add $j=0,\tfrac{M}{2}$ terms in the second and the forth terms of Eq.\eqref{Indexeven2}. Then, we find
\begin{align}
I_{\eta=+1}&=
\sum_{n=0}^{\infty} \sum_{j=0}^{\frac{M}{2}}
\psi_{T^2/{\mathbb{Z}}_2+,n}^{(j,0)}(z)_{+1}[\psi_{T^2/{\mathbb{Z}}_2+,n}^{(j,0)}(z)_{+1}]^{\ast} 
 -\sum_{n=1}^{\infty} \sum_{j=0}^{\frac{M}{2}}
\psi_{T^2/{\mathbb{Z}}_2-,n}^{(j,0)}(z)_{+1}[\psi_{T^2/{\mathbb{Z}}_2-,n}^{(j,0)}(z)_{+1}]^{\ast}
\notag \\
&=
\sum_{n=0}^{\infty} \sum_{j=0}^{\frac{M}{2}} \sum_{l=0}^{1} \sum_{l^{\prime}=0}^{1}
(1-\omega^{-l+l^{\prime}})|{\cal{N}}_{+,+1}^{(j,0)}|^2
\psi_{T^2+,n}^{(j,0)}(\omega^l z)[\psi_{T^2+,n}^{(j,0)}(\omega^{l^{\prime}}z)]^{\ast} \notag \\
&=\sum_{n=0}^{\infty} \sum_{j=0}^{\frac{M}{2}} 
(1-\omega)|{\cal{N}}_{+,+1}^{(j,0)}|^2
\Big{\{}\psi_{T^2+,n}^{(j,0)}(z)[\psi_{T^2+,n}^{(j,0)}(\omega z)]^{\ast}+
\psi_{T^2+,n}^{(j,0)}(\omega z)[\psi_{T^2+,n}^{(j,0)}( z)]^{\ast}\Big{\}} ,
\label{Indexeven3} 
\end{align}
where we have used the relations \eqref{Z2KKmode1}, \eqref{Z2KKmode} and ${\cal{N}}_{+,+1}^{(j,0)}={\cal{N}}_{+,-1}^{(j,0)}$ for $j=0,1,\cdots,\tfrac{M}{2}$ in the second equality, and $\omega=-1$ in the last equality.

Using the relation \eqref{Torusz2} and $\psi_{T^2+,n}^{(M,0)}(z)=\psi_{T^2+,n}^{(0,0)}(z)$, we can further rewrite Eq.\eqref{Indexeven3} as
\begin{align}
I_{\eta=+1}
&=\sum_{n=0}^{\infty}(1-\omega){\Big\{}2 |{\cal{N}}_{+,+1}^{(0,0)}|^2
\psi_{T^2+,n}^{(0,0)}(z)[\psi_{T^2+,n}^{(0,0)}(\omega z)]^{\ast}
+2 |{\cal{N}}_{+,+1}^{(M/2,0)}|^2
\psi_{T^2+,n}^{(M/2,0)}(z)[\psi_{T^2+,n}^{(M/2,0)}(\omega z)]^{\ast}
 \notag \\
&\quad+
 \sum_{j=1}^{\frac{M}{2}-1} |{\cal{N}}_{+,+1}^{(j,0)}|^2
(\psi_{T^2+,n}^{(j,0)}(z)[\psi_{T^2+,n}^{(j,0)}(\omega z)]^{\ast}+
\psi_{T^2+,n}^{(M-j,0)}( z)[\psi_{T^2+,n}^{(M-j,0)}( \omega z)]^{\ast}) {\Big\}}
\notag \\
&=\sum_{n=0}^{\infty} \sum_{j=0}^{M-1}(1-\omega)
\psi_{T^2+,n}^{(j,0)}(z)[\psi_{T^2+,n}^{(j,0)}(\omega z)]^{\ast},
\label{Indexeven4} 
\end{align}
where we have used ${\cal{N}}_{+,+1}^{(0,0)}={\cal{N}}_{+,+1}^{({M}/{2},0)}=1/\sqrt{2}$ and ${\cal{N}}_{+,+1}^{(j,0)}=1$ $(j=1,2,\cdots,\tfrac{M}{2}-1)$ for $M={\rm{even}}$ from Table \ref{T2}.

Taking the limit of $z^{\prime} \to z$ and $\rho \to \infty$ in Eq.\eqref{Indz2} and then substituting Eq.\eqref{Indexeven4} into Eq.\eqref{Indz2}, we find that Eq.\eqref{Ind3} holds for $(\alpha_1,\alpha_2)=(0,0),\,M={\rm{even}}$ and $\eta=+1$.

\subsubsection*{\underline{$(\alpha_1,\alpha_2)=(0,0),\,M=2m+2,\,\eta=-1$}}
We next discuss the case of $(\alpha_1,\alpha_2)=(0,0),\,M=2m+2$ and $\eta=-1$. Let us consider
\begin{align}
I_{\eta=-1}\equiv
\sum_{n=0}^{\infty} \sum_{j}
\psi_{T^2/{\mathbb{Z}}_2+,n}^{(j,0)}(z)_{-1}[\psi_{T^2/{\mathbb{Z}}_2+,n}^{(j,0)}(z)_{-1}]^{\ast}-\sum_{n=1}^{\infty} \sum_{j}
\psi_{T^2/{\mathbb{Z}}_2-,n}^{(j,0)}(z)_{-1}[\psi_{T^2/{\mathbb{Z}}_2-,n}^{(j,0)}(z)_{-1}]^{\ast}.
\label{Index2_1even}
\end{align}
From Table \ref{T1} and Eqs.\eqref{Z2KKmode1} and \eqref{Z2KKmode}, the label $j$ for $\psi_{T^2/{\mathbb{Z}}_2\pm,n}^{(j,0)}(z)_{-1}$ runs from 1 (0) to $\tfrac{M}{2}-1$ ($\tfrac{M}{2}$) for $n={\rm{even}}$ (odd). Thus, $I_{\eta=-1}$ can be expressed as
\begin{align}
I_{\eta=-1}=
&\sum_{k=0}^{\infty} \sum_{j=1}^{\frac{M}{2}-1}
\psi_{T^2/{\mathbb{Z}}_2+,2k}^{(j,0)}(z)_{-1}[\psi_{T^2/{\mathbb{Z}}_2+,2k}^{(j,0)}(z)_{-1}]^{\ast} \notag \\
+&\sum_{k=0}^{\infty} \sum_{j=0}^{\frac{M}{2}}
\psi_{T^2/{\mathbb{Z}}_2+,2k+1}^{(j,0)}(z)_{-1}[\psi_{T^2/{\mathbb{Z}}_2+,2k+1}^{(j,0)}(z)_{-1}]^{\ast} \notag \\
-&\sum_{k=0}^{\infty}\sum_{j=1}^{\frac{M}{2}-1}
\psi_{T^2/{\mathbb{Z}}_2-,2k+2}^{(j,0)}(z)_{-1}[\psi_{T^2/{\mathbb{Z}}_2-,2k+2}^{(j,0)}(z)_{-1}]^{\ast} \notag \\
-&\sum_{k=0}^{\infty} \sum_{j=0}^{\frac{M}{2}}
\psi_{T^2/{\mathbb{Z}}_2-,2k+1}^{(j,0)}(z)_{-1}[\psi_{T^2/{\mathbb{Z}}_2-,2k+1}^{(j,0)}(z)_{-1}]^{\ast}.
\label{Index2_2even}
\end{align}

Since $\psi_{T^2/{\mathbb{Z}}_2\pm,2k}^{(j,0)}(z)_{-1}=0$ for $j=0,\tfrac{M}{2}$, we can add $j=0,\tfrac{M}{2}$ terms in the first and the third terms of Eq.\eqref{Index2_2even}. Then, we find
\begin{align}
I_{\eta=-1}&=
\sum_{n=0}^{\infty} \sum_{j=0}^{\frac{M}{2}}
\psi_{T^2/{\mathbb{Z}}_2+,n}^{(j,0)}(z)_{-1}[\psi_{T^2/{\mathbb{Z}}_2+,n}^{(j,0)}(z)_{-1}]^{\ast} 
 -\sum_{n=1}^{\infty} \sum_{j=0}^{\frac{M}{2}}
\psi_{T^2/{\mathbb{Z}}_2-,n}^{(j,0)}(z)_{-1}[\psi_{T^2/{\mathbb{Z}}_2-,n}^{(j,0)}(z)_{-1}]^{\ast}
\notag \\
&=
\sum_{n=0}^{\infty} \sum_{j=0}^{\frac{M}{2}} \sum_{l=0}^{1} \sum_{l^{\prime}=0}^{1}\eta^{-l+l^{\prime}}(1{-}\omega^{-l+l^{\prime}})
|{\cal{N}}_{+,+1}^{(j,0)}|^2
\psi_{T^2+,n}^{(j,0)}(\omega^l z)[\psi_{T^2+,n}^{(j,0)}(\omega^{l^{\prime}}z)]^{\ast} \notag \\
&=\sum_{n=0}^{\infty} \sum_{j=0}^{\frac{M}{2}}
\eta(1-\omega) |{\cal{N}}_{+,+1}^{(j,0)}|^2
{\Big\{}\psi_{T^2+,n}^{(j,0)}(z)[\psi_{T^2+,n}^{(j,0)}(\omega z)]^{\ast}+
\psi_{T^2+,n}^{(j,0)}(\omega z)[\psi_{T^2+,n}^{(j,0)}( z)]^{\ast}{\Big\}} ,
\label{Index2_3even} 
\end{align}
where we have used the relations \eqref{Z2KKmode1}, \eqref{Z2KKmode} and ${\cal{N}}_{+,+1}^{(j,0)}={\cal{N}}_{+,-1}^{(j,0)}$ for $j=0,1,\cdots,\tfrac{M}{2}$ in the second equality, and $\eta=\omega=-1$ in the third equality.

Using the relation \eqref{Torusz2} and $\psi_{T^2+,n}^{(M,0)}(z)=\psi_{T^2+,n}^{(0,0)}(z)$, we can further rewrite Eq.\eqref{Index2_3even} as
\begin{align}
I_{\eta=-1}
&=\sum_{n=0}^{\infty}\eta(1-\omega){\Bigl\{}2 |{\cal{N}}_{+,+1}^{(0,0)}|^2
\psi_{T^2+,n}^{(0,0)}(z)[\psi_{T^2+,n}^{(0,0)}(\omega z)]^{\ast}
+2 |{\cal{N}}_{+,+1}^{(M/2,0)}|^2
\psi_{T^2+,n}^{(M/2,0)}(z)[\psi_{T^2+,n}^{(M/2,0)}(\omega z)]^{\ast}
 \notag \\
&\quad+
 \sum_{j=1}^{\frac{M}{2}-1} |{\cal{N}}_{+,+1}^{(j,0)}|^2
(\psi_{T^2+,n}^{(j,0)}(z)[\psi_{T^2+,n}^{(j,0)}(\omega z)]^{\ast}+
\psi_{T^2+,n}^{(M-j,0)}( z)[\psi_{T^2+,n}^{(M-j,0)}( \omega z)]^{\ast}) {\Bigr\}}
\notag \\
&=\sum_{n=0}^{\infty} \sum_{j=0}^{M-1}\eta(1-\omega)
\psi_{T^2+,n}^{(j,0)}(z)[\psi_{T^2+,n}^{(j,0)}(\omega z)]^{\ast},
\label{Index2_4even} 
\end{align}
where we have used ${\cal{N}}_{+,+1}^{(0,0)}={\cal{N}}_{+,+1}^{({M}/{2},0)}=1/\sqrt{2}$ and ${\cal{N}}_{+,+1}^{(j,0)}=1$ $(j=1,2,\cdots,\tfrac{M}{2}-1)$ for $M={\rm{even}}$ from Table \ref{T2}.

Taking the limit of $z^{\prime} \to z$ and $\rho \to \infty$ in Eq.\eqref{Indz2} and then substituting Eq.\eqref{Index2_4even} into Eq.\eqref{Indz2}, we find that Eq.\eqref{Ind3} holds for $(\alpha_1,\alpha_2)=(0,0),\,M={\rm{even}}$ and $\eta=-1$.
We have succeeded in verifying Eq.\eqref{Ind3} for the case of $(\alpha_1,\alpha_2)=(0,0),\,M={\rm{even}}$ and $\eta=\pm 1$.
\color{black}
 We can similarly show that Eq.\eqref{Ind3} holds for other cases of $(\alpha_1,\alpha_2),\,M$ and $\eta$.

\section{{Derivation of the formula \eqref{WDderive}}}
In this Appendix, we derive the formula \eqref{WDderive} for the $T^2/{\mathbb{Z}}_N$ $(N=2,3,4,6)$ orbifolds.

\subsection{$T^2/{\mathbb{Z}}_2$}
From Eqs.\eqref{Inddelta2} and \eqref{IndWp}, we have
\begin{align}
W_p&={\eta e^{i \theta_{m,n}(z_p^f)}} \notag \\
&=
\eta e^{imq\Lambda_1(z_p^f)+i2\pi m\alpha_1+inq\Lambda_2(z_p^f)+i2\pi n\alpha_2+i\pi M mn}\notag \\
&=\omega^{\chi_{+p}} \qquad(p=1,2,3,4),
\end{align}
where we have used $\eta=\omega^l\,\,(l=0,1)$ and Eqs.\eqref{thetamn} and \eqref{WDN1}. Since $\chi_{+p}$ is taken to be 0 or 1, the following identity holds:
\begin{align}
\omega^{\chi_{+p}}=-2\chi_{+p}+1 \qquad(\omega=-1).
\end{align}
Thus, we obtain
\begin{align}
W_p=-2\chi_{+p}+1,
\end{align}
which is the relation \eqref{WDderive} for $T^2/{\mathbb{Z}}_2$.

\subsection{$T^2/{\mathbb{Z}}_3$}
From Eqs.\eqref{IndW31}$-$\eqref{IndW33}, we find
\begin{align}
W_p&=\frac{2}{3}\{\eta(1-\omega)e^{i\theta_{m_1,n_1}(z_p^f)}+\eta^2(1-\omega^2)e^{i\theta_{m_2,n_2}(z_p^f)}\} \notag \\
&=\frac{2}{3}\{\eta(1-\omega)e^{i\theta_{m_1,n_1}(z_p^f)}+\eta^2(1-\omega^2)e^{i2\theta_{m_1,n_1}(z_p^f)}\} \notag \\
&=\frac{2}{3}\{(1-\omega)\omega^{\chi_{+p}}+(1-\omega^2)\omega^{2\chi_{+p}}\},
\label{Z3AW}
\end{align}
where $(m_k,n_k)$ are determined from the fixed point equations $z_p^f=\omega^k z_p^f+m_k+n_k \tau\,\,({k=1,2})$. At the second equality of Eq.\eqref{Z3AW}, we have used the relation
\begin{align}
e^{i\theta_{m_2,n_2}(z_p^f)}=e^{i2\theta_{m_1,n_1}(z_p^f)}.
\label{A5}
\end{align}
The proof of Eq.\eqref{A5} will be given in the subsection A.5.

For $\chi_{+p}=0,1,2$, the following identity holds
\begin{align}
\omega^{\chi_{+p}}=1+\left(\frac{5}{2}\omega-1\right)\chi_{+p}-\frac{3}{2}\omega{\chi^2_{+p}}
\label{Z3omega}
\end{align}
with $\omega=e^{i2\pi/3}$. Inserting Eq.\eqref{Z3omega} into Eq.\eqref{Z3AW} and using $\chi_{+p}(\chi_{+p}-1)(\chi_{+p}-2)=0$, we have
\begin{align}
W_p=-2\chi_{+p}+2.
\end{align}

\subsection{$T^2/{\mathbb{Z}}_4$}
We first note that the $T^2/{\mathbb{Z}}_4$ orbifold has two ${\mathbb{Z}}_4$ fixed points ($z_1^f=0,\,z_2^f=(1+\tau)/2)$ and two ${\mathbb{Z}}_2$ fixed points ($z_3^f=1/2,\,z_4^f=\tau/2)$. For the ${\mathbb{Z}}_4$ fixed points $z_p^f\,\,(p=1,2)$, from Eqs.\eqref{IndW41} and \eqref{IndW42} we obtain
\begin{align}
W_p&=\eta(1-\omega)e^{i\theta_{m_1,n_1}(z_p^f)}+\frac{1}{2}\eta^2(1-\omega^2)e^{i\theta_{m_2,n_2}(z_p^f)}+\eta^3(1-\omega^3)e^{i\theta_{m_3,n_3}(z_p^f)} \notag \\
&=\eta(1-\omega)e^{i\theta_{m_1,n_1}(z_p^f)}+\frac{1}{2}\eta^2(1-\omega^2)e^{i2\theta_{m_1,n_1}(z_p^f)}+\eta^3(1-\omega^3)e^{i3\theta_{m_1,n_1}(z_p^f)} \notag \\
 &=(1-\omega)\omega^{\chi_{+p}}+\frac{1}{2}(1-\omega^2)\omega^{2\chi_{+p}}+(1-\omega^3)\omega^{3\chi_{+p}},
\label{Z4AW}
\end{align}
where we have used
\begin{align}
e^{i\theta_{m_k,n_k}(z_p^f)}=e^{ik\theta_{m_1,n_1}(z_p^f)} \qquad(k=2,3),
\label{A9}
\end{align}
and $(m_k,n_k)$ are solutions to the fixed point equations $z_p^f=\omega^k z_p^f+m_k+n_k \tau\,\,(k=1,2,3)$. 

Using the identities
\begin{gather}
\omega^{\chi_{+p}}=1+\left(-\frac{1}{3}+\frac{8}{3}\omega\right)\chi_{+p}-(1+2\omega){\chi^2_{+p}}+\left(\frac{1}{3}+\frac{1}{3}\omega\right)\chi^3_{+p},
\\
\chi_{+p}(\chi_{+p}-1)(\chi_{+p}-2)(\chi_{+p}-3)=0,
\label{Z4omega}
\end{gather}
for $\chi_{+p}=0,1,2,3$ with $\omega=e^{i2\pi/4}=i$, we can show that the right-hand side of Eq.\eqref{Z4AW} reduces to
\begin{align}
W_p=-2\chi_{+p}+3 \qquad {\rm{for}} \quad p=1,2.
\end{align}

For the ${\mathbb{Z}}_2$ fixed point $z_3^f=1/2$ with $z_3^f=\omega^2 z_3^f+m_2+n_2\tau$, from Eq.\eqref{IndW43} we have
\begin{align}
W_3&=\frac{1}{2}\eta^2(1-\omega^2)e^{i\theta_{m_2,n_2}(z_3^f)} \notag \\
&={(\omega^{2})^{\chi_{+3}}} \notag \\
&=-2\chi_{+3}+1,
\label{A13}
\end{align}
where we have used $\omega^2=-1$, $\eta^2=(\omega^2)^l\,\,(l=0,1)$ and Eq.\eqref{WDN1} with $N=2$. In the last equality of Eq.\eqref{A13}, we have used the analysis in the subsection A.1. A similar discussion holds for another ${\mathbb{Z}}_2$ fixed point $z_4^f=\tau/2$, and we find
\begin{align}
W_4=-2\chi_{+4}+1.
\label{A14}
\end{align}

\subsection{$T^2/{\mathbb{Z}}_6$}
We first note that the $T^2/{\mathbb{Z}}_6$ orbifold contains
\begin{align}
&{\rm{one \,\,}}{\mathbb{Z}}_6{\rm{\,\,fixed \,\,point\,\,}}: z_1^f=0, \notag \\
&{\rm{two \,\,}}{\mathbb{Z}}_3{\rm{\,\,fixed \,\,points\,\,}}: z_2^f=(1+\tau)/3,\,\, z_3^f=2(1+\tau)/3, \notag \\
&{\rm{three \,\,}}{\mathbb{Z}}_2{\rm{\,\,fixed \,\,points\,\,}}: z_4^f=1/2\,\,z_5^f=\tau/2\,\, z_6^f=(1+\tau)/2.
\end{align}
For the ${\mathbb{Z}}_6$ fixed point $z_1^f$, from \eqref{IndW61} we have
\begin{align}
W_1&=2\eta(1-\omega)e^{i\theta_{m_1,n_1}(z_1^f)}+\frac{2}{3}\eta^2(1-\omega^2)e^{i\theta_{m_2,n_2}(z_1^f)}+\frac{1}{2}\eta^3(1-\omega^3)e^{i\theta_{m_3,n_3}(z_1^f)}  \notag \\
&\qquad \qquad +\frac{2}{3}\eta^4(1-\omega^4)e^{i\theta_{m_4,n_4}(z_1^f)}
+2\eta^5(1-\omega^5)e^{i\theta_{m_5,n_5}(z_1^f)}\notag \\
&=2(1-\omega)\omega^{\chi_{+1}}+\frac{2}{3}(1-\omega^2)\omega^{2\chi_{+1}}+\frac{1}{2}(1-\omega^3)\omega^{3\chi_{+1}}  \notag \\
&\qquad \qquad+\frac{2}{3}(1-\omega^4)\omega^{4\chi_{+1}}
+2(1-\omega^5)\omega^{5\chi_{+1}},
\label{Z6AW}
\end{align}
we have used the relations
\begin{align}
e^{i\theta_{m_k,n_k}(z_1^f)}=e^{ik\theta_{m_1,n_1}(z_1^f)} \qquad(k=1,2,3,4,5),
\label{A17}
\end{align}
and $\eta e^{i\theta_{m_1,n_1}(z_1^f)}=\omega^{\chi_{+1}}$.

Using the identities
\begin{align}
\omega^{\chi_{+1}}&=1+\frac{1}{60}(-25+63\omega)\chi_{+1}+\frac{1}{24}(-23+9\omega){\chi^2_{+1}} \notag \\
&\qquad+\frac{1}{24}(10-13\omega)\chi^3_{+1}+\frac{1}{24}(-1+3\omega)\chi^4_{+1}
+\frac{1}{120}(-\omega)\chi^5_{+1}, \label{A18}\\
\chi_{+1}&(\chi_{+1}-1)(\chi_{+1}-2)(\chi_{+1}-3)(\chi_{+1}-4)(\chi_{+1}-5)=0
\label{constrainZ6}
\end{align}
with $\chi_{+1}=0,1,\cdots,5$ and $\omega=e^{i2\pi/6}$, we can show that Eq.\eqref{Z6AW} takes the form
\begin{align}
W_1=-2\chi_{+1}+5.
\end{align}

For the ${\mathbb{Z}}_3$ fixed point $z_2^f=(1+\tau)/3$ with $z_2^f=\omega^2 z_2^f+m_2+n_2\tau$ and $z_2^f=\omega^4 z_2^f+m_4+n_4\tau$, from Eq.\eqref{IndW62} we have
\begin{align}
W_2&=\frac{2}{3}\eta^2(1-\omega^2)e^{i\theta_{m_2,n_2}(z_2^f)}
+\frac{2}{3}\eta^4(1-\omega^4)e^{i\theta_{m_4,n_4}(z_2^f)} \notag \\
&=\frac{2}{3}(1-\omega^2){({\omega}^2)}^{\chi_{+2}}+\frac{2}{3}(1-\omega^4){({\omega}^2)}^{2\chi_{+2}} \notag \\
&=-2\chi_{+2}+2,
\label{A21}
\end{align}
where we have used Eq.\eqref{WDN1} with $N=3$ and $\eta^2=(\omega^2)^l\,\,(l=0,1,2)$. In the last equality of Eq.\eqref{A21}, we followed the analysis in the subsection A.2. A similar analysis for another ${\mathbb{Z}}_3$ fixed point $z_3^f=2(1+\tau)/3$ shows
\begin{align}
W_3=-2\chi_{+3}+2.
\label{A22}
\end{align}

For the ${\mathbb{Z}}_2$ fixed point $z_4^f=1/2$, $z_5^f=\tau/2$ and $z_6^f=(1+\tau)/2$ with $z_p^f=\omega^3 z_p^f+m_3+n_3\tau\,\,(p=4,5,6)$, from Eqs.\eqref{IndW64}$-$\eqref{IndW66} we have
\begin{align}
W_p&=\frac{1}{2}\eta^3(1-\omega^3)e^{i\theta_{m_3,n_3}(z_p^f)}
 \notag \\
&=\frac{1}{2}(1-\omega^3){({\omega}^3)}^{\chi_{+p}}\notag \\
&=-2\chi_{+p}+1\qquad(p=4,5,6),
\label{A23}
\end{align}
where we have used Eq.\eqref{WDN1} with $N=2$ and $\eta^3=(\omega^3)^l\,\,(l=0,1)$. In the last equality of Eq.\eqref{A23}, we followed the analysis in the subsection A.1.

\subsection{Proof of Eqs.\eqref{A5}, \eqref{A9} and \eqref{A17}}
In this subsection, we prove Eqs.\eqref{A5}, \eqref{A9} and \eqref{A17}. To this end, we start with the relation \eqref{WDdefine}
\begin{align}
\psi_{T^2/{\mathbb{Z}}_N+,n}^{(j+\alpha_1,\alpha_2)}(z_p^f+\omega Z)_{\eta}&=e^{i(m_1q\Lambda_1(\omega Z)+n_1q\Lambda_2(\omega Z))}e^{i\theta_{m_1,n_1}(z_p^f)}\eta \, \psi_{T^2/{\mathbb{Z}}_N+,n}^{(j+\alpha_1,\alpha_2)}(z_p^f+ Z)_{\eta},
\label{A241}
\end{align}
where $m_1$ and $n_1$ are defined through the fixed point equation
\begin{align}
z_p^f=\omega z_p^f+m_1+n_1 \tau
\label{A24}
\end{align}
and we have {used} the definition \eqref{thetamn} of $\theta_{m,n}(z)$. 

In the following, we first show the relation
\begin{align}
e^{i\theta_{m_2,n_2}(z_p^f)}=e^{i2\theta_{m_1,n_1}(z_p^f)} ,
\label{A25}
\end{align}
where $(m_1,n_1)$ and $(m_2,n_2)$ are related to $z_p^f$ as
\begin{align}
z_p^f&=\omega z_p^f+m_1+n_1 \tau, \label{A26}\\
z_p^f&=\omega^2 z_p^f+m_2+n_2 \tau,
\label{A27}
\end{align}
by computing $\psi_{T^2/{\mathbb{Z}}_N+,n}^{(j+\alpha_1,\alpha_2)}(z_p^f+\omega^2 Z)_{\eta}$ in two ways.

{We} first evaluate $\psi_{T^2/{\mathbb{Z}}_N+,n}^{(j+\alpha_1,\alpha_2)}(z_p^f+\omega^2 Z)_{\eta}$, by use of the formula \eqref{A241} twice, as follows:
\begin{align}
\psi_{T^2/{\mathbb{Z}}_N+,n}^{(j+\alpha_1,\alpha_2)}(z_p^f+\omega^2 Z)_{\eta}
&=\psi_{T^2/{\mathbb{Z}}_N+,n}^{(j+\alpha_1,\alpha_2)}(z_p^f+\omega(\omega Z))_{\eta} \notag \\
&=e^{im_1q\Lambda_1(\omega^2 Z)+in_1q\Lambda_2(\omega^2 Z)}e^{i\theta_{m_1,n_1}(z_p^f)}\eta\psi_{T^2/{\mathbb{Z}}_N+,n}^{(j+\alpha_1,\alpha_2)}(z_p^f+\omega Z)_{\eta} \notag \\
&=e^{im_1q\Lambda_1(\omega^2 Z)+in_1q\Lambda_2(\omega^2 Z)}
e^{im_1q\Lambda_1(\omega Z)+in_1q\Lambda_2(\omega Z)} \notag \\
&\qquad \times e^{i2\theta_{m_1,n_1}(z_p^f)}\eta^2 \psi_{T^2/{\mathbb{Z}}_N+,n}^{(j+\alpha_1,\alpha_2)}(z_p^f+ Z)_{\eta}.
\label{A28}
\end{align}
Next, we evaluate $\psi_{T^2/{\mathbb{Z}}_N+,n}^{(j+\alpha_1,\alpha_2)}(z_p^f+\omega^2 Z)_{\eta}$ with Eq.\eqref{A27} as follows:
\begin{align}
\psi_{T^2/{\mathbb{Z}}_N+,n}^{(j+\alpha_1,\alpha_2)}(z_p^f+\omega^2 Z)_{\eta}
&=\psi_{T^2/{\mathbb{Z}}_N+,n}^{(j+\alpha_1,\alpha_2)}(\omega^2 z_p^f+m_2+n_2 \tau+\omega^2  Z)_{\eta} \notag \\
&=e^{im_2q\Lambda_1(\omega^2 Z)+in_2q\Lambda_2(\omega^2 Z)}e^{i\theta_{m_2,n_2}(z_p^f)}\eta^2\psi_{T^2/{\mathbb{Z}}_N+,n}^{(j+\alpha_1,\alpha_2)}(z_p^f+ Z)_{\eta},
\label{A29}
\end{align}
where we have used the boundary conditions \eqref{BCs} and the $\mathbb{Z}_N$ transformation \eqref{ZNes1}. Equating Eq.\eqref{A28} with Eq.{\eqref{A29} and taking the limit of $Z \to 0$, we have 
\begin{align}
e^{i\theta_{m_2,n_2}(z_p^f)}=e^{i2\theta_{m_1,n_1}(z_p^f)} .
\label{A30}
\end{align}

Similarly, for the fixed point $z_p^f$ which satisfies the fixed point equations
\begin{align}
z_p^f&=\omega z_p^f+m_1+n_1 \tau, \\
z_p^f&=\omega^k z_p^f+m_k+n_k \tau,
\label{A32}
\end{align}
we can show 
\begin{align}
e^{i\theta_{m_k,n_k}(z_p^f)}=e^{ik\theta_{m_1,n_1}(z_p^f)} .
\label{A33}
\end{align}

\bibliographystyle{unsrt}
 \bibliography{referenceindex}

\begin{thebibliography}{10}

\bibitem{Sakamoto:2020pev}
Makoto Sakamoto, Maki Takeuchi, and Yoshiyuki Tatsuta.
\newblock {Zero-mode counting formula and zeros in orbifold compactifications}.
\newblock {\em Phys. Rev. D}, 102(2):025008, 2020.

\bibitem{Abe:2008sx}
Hiroyuki Abe, Kang-Sin Choi, Tatsuo Kobayashi, and Hiroshi Ohki.
\newblock {Three generation magnetized orbifold models}.
\newblock {\em Nucl. Phys. B}, 814:265--292, 2009.

\bibitem{Abe:2015yva}
Tomo-hiro Abe, Yukihiro Fujimoto, Tatsuo Kobayashi, Takashi Miura, Kenji
  Nishiwaki, Makoto Sakamoto, and Yoshiyuki Tatsuta.
\newblock {Classification of three-generation models on magnetized orbifolds}.
\newblock {\em Nucl. Phys. B}, 894:374--406, 2015.

\bibitem{Libanov:2000uf}
M.~V. Libanov and Sergey~V. Troitsky.
\newblock {Three fermionic generations on a topological defect in extra
  dimensions}.
\newblock {\em Nucl. Phys. B}, 599:319--333, 2001.

\bibitem{Frere:2000dc}
J.~M. Frere, M.~V. Libanov, and Sergey~V. Troitsky.
\newblock {Three generations on a local vortex in extra dimensions}.
\newblock {\em Phys. Lett. B}, 512:169--173, 2001.

\bibitem{PhysRevD.65.044004}
Andrey Neronov.
\newblock Fermion masses and quantum numbers from extra dimensions.
\newblock {\em Phys. Rev. D}, 65:044004, Jan 2002.

\bibitem{PhysRevD.73.085007}
Silvestre Aguilar and Douglas Singleton.
\newblock Fermion generations, masses, and mixings in a 6d brane model.
\newblock {\em Phys. Rev. D}, 73:085007, Apr 2006.

\bibitem{Gogberashvili:2007gg}
Merab Gogberashvili, Pavle Midodashvili, and Douglas Singleton.
\newblock {Fermion Generations from 'Apple-Shaped' Extra Dimensions}.
\newblock {\em JHEP}, 08:033, 2007.

\bibitem{Guo:2008ia}
Zhi-qiang Guo and Bo-Qiang Ma.
\newblock {Fermion Families from Two Layer Warped Extra Dimensions}.
\newblock {\em JHEP}, 08:065, 2008.

\bibitem{PhysRevLett.108.181807}
David~B. Kaplan and Sichun Sun.
\newblock Spacetime as a topological insulator: Mechanism for the origin of the
  fermion generations.
\newblock {\em Phys. Rev. Lett.}, 108:181807, May 2012.

\bibitem{Cremades:2004wa}
D.~Cremades, L.~E. Ibanez, and F.~Marchesano.
\newblock {Computing Yukawa couplings from magnetized extra dimensions}.
\newblock {\em JHEP}, 05:079, 2004.

\bibitem{Arkani-Hamed:1999ylh}
Nima Arkani-Hamed and Martin Schmaltz.
\newblock {Hierarchies without symmetries from extra dimensions}.
\newblock {\em Phys. Rev. D}, 61:033005, 2000.

\bibitem{Dvali:2000ha}
G.~R. Dvali and Mikhail~A. Shifman.
\newblock {Families as neighbors in extra dimension}.
\newblock {\em Phys. Lett. B}, 475:295--302, 2000.

\bibitem{Gherghetta:2000qt}
Tony Gherghetta and Alex Pomarol.
\newblock {Bulk fields and supersymmetry in a slice of AdS}.
\newblock {\em Nucl. Phys. B}, 586:141--162, 2000.

\bibitem{Kaplan:2000av}
David~Elazzar Kaplan and Timothy M.~P. Tait.
\newblock {Supersymmetry breaking, fermion masses and a small extra dimension}.
\newblock {\em JHEP}, 06:020, 2000.

\bibitem{Huber:2000ie}
Stephan~J. Huber and Qaisar Shafi.
\newblock {Fermion masses, mixings and proton decay in a Randall-Sundrum
  model}.
\newblock {\em Phys. Lett. B}, 498:256--262, 2001.

\bibitem{Kaplan:2001ga}
David~Elazzar Kaplan and Timothy M.~P. Tait.
\newblock {New tools for fermion masses from extra dimensions}.
\newblock {\em JHEP}, 11:051, 2001.

\bibitem{Fujimoto:2012wv}
Yukihiro Fujimoto, Tomoaki Nagasawa, Kenji Nishiwaki, and Makoto Sakamoto.
\newblock {Quark mass hierarchy and mixing via geometry of extra dimension with
  point interactions}.
\newblock {\em PTEP}, 2013:023B07, 2013.

\bibitem{PhysRevD.97.115039}
Yukihiro Fujimoto, Takashi Miura, Kenji Nishiwaki, and Makoto Sakamoto.
\newblock Dynamical generation of fermion mass hierarchy in an extra dimension.
\newblock {\em Phys. Rev. D}, 97:115039, Jun 2018.

\bibitem{PhysRevD.90.105006}
Hiroyuki Abe, Tatsuo Kobayashi, Keigo Sumita, and Yoshiyuki Tatsuta.
\newblock Gaussian froggatt-nielsen mechanism on magnetized orbifolds.
\newblock {\em Phys. Rev. D}, 90:105006, Nov 2014.

\bibitem{PhysRevD.88.115007}
Yukihiro Fujimoto, Kenji Nishiwaki, and Makoto Sakamoto.
\newblock $cp$ phase from twisted higgs vacuum expectation value in extra
  dimension.
\newblock {\em Phys. Rev. D}, 88:115007, Dec 2013.

\bibitem{Kobayashi:2016qag}
Tatsuo Kobayashi, Kenji Nishiwaki, and Yoshiyuki Tatsuta.
\newblock {CP-violating phase on magnetized toroidal orbifolds}.
\newblock {\em JHEP}, 04:080, 2017.

\bibitem{Buchmuller:2017vho}
Wilfried Buchmuller and Julian Schweizer.
\newblock {Flavor mixings in flux compactifications}.
\newblock {\em Phys. Rev. D}, 95(7):075024, 2017.

\bibitem{PhysRevD.97.075019}
Wilfried Buchmuller and Ketan~M. Patel.
\newblock Flavor physics without flavor symmetries.
\newblock {\em Phys. Rev. D}, 97:075019, Apr 2018.

\bibitem{Dixon:1985jw}
Lance~J. Dixon, Jeffrey~A. Harvey, C.~Vafa, and Edward Witten.
\newblock {Strings on Orbifolds}.
\newblock {\em Nucl. Phys. B}, 261:678--686, 1985.

\bibitem{Dixon:1986jc}
Lance~J. Dixon, Jeffrey~A. Harvey, C.~Vafa, and Edward Witten.
\newblock {Strings on Orbifolds. 2.}
\newblock {\em Nucl. Phys. B}, 274:285--314, 1986.

\bibitem{Atiyah:1963zz}
M.~F. Atiyah and I.~M. Singer.
\newblock {The index of elliptic operators on compact manifolds}.
\newblock {\em Bull. Am. Math. Soc.}, 69:422--433, 1969.

\bibitem{Witten:1984dg}
Edward Witten.
\newblock {Some Properties of O(32) Superstrings}.
\newblock {\em Phys. Lett. B}, 149:351--356, 1984.

\bibitem{Green:1987mn}
Michael~B. Green, J.~H. Schwarz, and Edward Witten.
\newblock {\em {SUPERSTRING THEORY. VOL. 2: LOOP AMPLITUDES, ANOMALIES AND
  PHENOMENOLOGY}}.
\newblock 7 1988.

\bibitem{Abe:2013bca}
Tomo-Hiro Abe, Yukihiro Fujimoto, Tatsuo Kobayashi, Takashi Miura, Kenji
  Nishiwaki, and Makoto Sakamoto.
\newblock {$Z_N$ twisted orbifold models with magnetic flux}.
\newblock {\em JHEP}, 01:065, 2014.

\bibitem{Abe:2014noa}
Tomo-hiro Abe, Yukihiro Fujimoto, Tatsuo Kobayashi, Takashi Miura, Kenji
  Nishiwaki, and Makoto Sakamoto.
\newblock {Operator analysis of physical states on magnetized $T^{2}/Z_{N}$
  orbifolds}.
\newblock {\em Nucl. Phys. B}, 890:442--480, 2014.

\bibitem{PhysRevD.96.096011}
Tatsuo Kobayashi and Satoshi Nagamoto.
\newblock Zero-modes on orbifolds: Magnetized orbifold models by modular
  transformation.
\newblock {\em Phys. Rev. D}, 96:096011, Nov 2017.

\bibitem{Sakamoto:2020vdy}
Makoto Sakamoto, Maki Takeuchi, and Yoshiyuki Tatsuta.
\newblock {Index theorem on $T^2/\mathbb{Z}_N$ orbifolds}.
\newblock {\em Phys. Rev. D}, 103(2):025009, 2021.

\bibitem{PhysRevLett.42.1195}
Kazuo Fujikawa.
\newblock Path-integral measure for gauge-invariant fermion theories.
\newblock {\em Phys. Rev. Lett.}, 42:1195--1198, Apr 1979.

\bibitem{PhysRevD.22.1499}
Kazuo Fujikawa.
\newblock Erratum: Path integral for gauge theories with fermions.
\newblock {\em Phys. Rev. D}, 22:1499--1499, Sep 1980.

\bibitem{Kobayashi:2022tti}
Tatsuo Kobayashi, Hajime Otsuka, Makoto Sakamoto, Maki Takeuchi, Yoshiyuki
  Tatsuta, and Hikaru Uchida.
\newblock {Index theorem on magnetized blow-up manifold of $T^2/\mathbb{Z}_N$}.
\newblock 11 2022.

\bibitem{Kobayashi:2022xsk}
Tatsuo Kobayashi, Hajime Otsuka, Makoto Sakamoto, Maki Takeuchi, Yoshiyuki
  Tatsuta, and Hikaru Uchida.
\newblock {Zero-mode wave functions by localized gauge fluxes}.
\newblock 11 2022.

\end{thebibliography}

\end{document}